\documentclass[journal,onecolumn]{IEEEtran}

\usepackage{graphicx,amsthm,amsmath,latexsym,amssymb,times,bbold}
\usepackage{float,epsfig,multirow,rotating,times,verbatim,wrapfig,booktabs}
\usepackage{color,array,hyperref, algpseudocode}
\usepackage{cite}

\usepackage{array}
\newenvironment{strip}
{}
{}

\newtheorem{thm}{Theorem}[section]

\newtheorem{claim}[thm]{Claim}
\newcommand{\citep}{\cite}
\newcommand{\citet}{\cite}
\newcommand{\citeyear}{\cite}
\newcommand{\citeauthor}{\cite}

\usepackage[obeyFinal,colorinlistoftodos]{todonotes}

\usepackage{standalone}

\usepackage{float}

\usepackage{amsfonts,amsopn}
\usepackage{bbold}
\usepackage{bm}
\usepackage{comment}
\usepackage{multirow}
\usepackage{color,soul}
\usepackage{xr-hyper}
\usepackage{hyperref}
\usepackage{rotating}
\usepackage[caption=false,font=footnotesize]{subfig}

\usepackage{booktabs}
\usepackage{siunitx}

\usepackage{etoolbox}	
\renewcommand{\bfseries}{\fontseries{b}\selectfont}
\robustify\bfseries 
\newrobustcmd{\B}{\bfseries}
\newrobustcmd{\Tnote}[1]{\textsuperscript {\TPTtagStyle {#1}}}	

\usepackage{threeparttable}

\usepackage{tikz}
\usetikzlibrary{arrows.meta,bending,trees,snakes}
\usepackage{pgf}

\newcommand{\hw}{OTQT}
\newcommand{\hwo}{OT}

\newcommand{\figref}{Fig.~}

\newcommand{\bem}{\ensuremath{\bm{m}}}

\newcommand{\bX}{\ensuremath{\bm{X}}}

\newcommand{\mC}{\ensuremath{\mathcal{C}}}
\newcommand{\mJ}{\ensuremath{\mathcal{J}}}
\newcommand{\mL}{\ensuremath{\mathcal{L}}}
\newcommand{\mW}{\ensuremath{\mathcal{W}}}
\newcommand{\mbC}{\ensuremath{\bm{\mathcal{C}}}}
\newcommand{\mbX}{\ensuremath{\bm{\mathcal{X}}}}

\newcommand{\bmu}{{\ensuremath{\bm{\mu}}}}

\newcommand{\bpi}{{\ensuremath{\bm{\pi}}}}

\newcommand{\1}{\mathbb{1}}

\DeclareMathOperator{\E}{\mathbb{E}}
\DeclareMathOperator*{\argmax}{arg\,max}
\newcommand{\argmin}{\operatornamewithlimits{argmin}}

\usepackage{soul}

\newcommand{\del}[1]{}			

\begin{document}
\title{An Efficient $k$-modes Algorithm for Clustering Categorical Datasets}
\author{Karin~S.~Dorman~and~Ranjan~Maitra

  \thanks{K. S. Dorman is with the Departments of Statistics and Genetics Development and Cell Biology at Iowa State
    University, Ames, Iowa, USA.}
  \thanks{R. Maitra is with the Department of Statistics, Iowa State
    University, Ames, Iowa, USA.}
  \thanks{This research was supported in part by the United States    Department of Agriculture (USDA) National Institute of Food and Agriculture (NIFA) Hatch project IOW03617. The content of this paper however is solely the responsibility of the   authors and does not represent the official views of either the NIFA or the USDA.}

}

\markboth{}%
{Almod\'ovar-Rivera and Maitra: Nonparametric Syncytial Clustering}


\IEEEcompsoctitleabstractindextext{%

    Mining clusters from data is an important endeavor in many applications.
The $k$-means method is a popular, efficient, and distribution-free approach for clustering numerical-valued data, but does not apply for categorical-valued observations.
The $k$-modes method addresses this lacuna by replacing the Euclidean with the Hamming distance and the means with the modes in the $k$-means objective function.
We provide a novel, computationally efficient implementation of $k$-modes, called \hw.
We prove that \hw{} finds updates to improve the objective function that are undetectable to existing $k$-modes algorithms. 
Although slightly slower per iteration due to algorithmic complexity, \hw{} is always more accurate per iteration and almost always faster (and only barely slower on some datasets) to the final optimum.
Thus, we recommend \hw{} as the preferred, default algorithm for $k$-modes optimization.

  \begin{IEEEkeywords}
categorical data clustering,
$k$-modes, OT algorithm, OTQT algorithm
\end{IEEEkeywords}}
\maketitle

\IEEEdisplaynotcompsoctitleabstractindextext

\section{Introduction}
Identifying groups of similar observations in datasets is common in a wide array of applications, with many clustering methods developed in statistics, machine learning and the applied sciences~\citep{ramey85, mclachlanandbasford88, kaufmanandrousseuw90, everittetal01, melnykovandmaitra10, xuandwunsch09, Bouveyron2019}. 
The $k$-means algorithm~\citep{Jancey1966,macqueen67,lloyd82,hartiganandwong79} is arguably the most popular method for clustering numerical-valued observations.
It scales to large datasets because it does not require calculation of
all pairwise distances. 
The method is also distribution-free, which while not equivalent to 
assumption-free~\citep{celeuxandgovaert92,maitraetal12}, makes it a popular starting point for users wary of imposing distributional assumptions on their data.
Unfortunately, the $k$-means objective function is inapplicable to data with categorical attributes.

There exist several methods for clustering categorical-featured datasets~\citep{Andreopoulos2014}, 
but the most direct counterpart to $k$-means is
$k$-modes
\footnote{Another clustering algorithm with the same name~\citep{Carreira-Perpinan2013} is intended for continuous data.}~\citep{Chaturvedi2001,huang97b},
where each cluster is 
characterized by its sample mode, rather than its sample mean, and the objective
function is the sum of distances, usually Hamming distances, between the observations and their
respective cluster modes. 
The $k$-modes objective function is most commonly minimized
    through the \citet{huang97b} algorithm that
  mimics the~\citet{macqueen67} $k$-means
  algorithm~\citep{Selim1984}.
It is initialized with $k$ modes, and then alternately (a) allocates the next observation to the closest mode and (b) updates the affected modes, continuing to cycle through the observations until no change occurs during a complete pass.

There have been variations proposed on MacQueen's $k$-means algorithm, with some \del{being }provably superior\del{ algorithms}.
Lloyd's algorithm~\citep{lloyd82} updates the means only \textit{after all} the observations have been reallocated to the clusters, but there appears to be little practical difference with MacQueen's algorithm~\citep{HarPeled2005, Telgarsky2010}.
On the other hand, \citet{hartigan75} shifts an observation between clusters not based on distance to the means, but in order to achieve the biggest reduction in the objective function, a change proven to achieve better optima~\citep{Telgarsky2010}.
\citet{hartiganandwong79} make further efficiency tweaks to avoid attempts to move observations to clusters that have not changed since the last pass through the data and a ``quick-transfer stage,'' where observations are only swapped between the two closest clusters. 
While its intended purpose is not proven, the quick-transfer stage may
offer suboptimal, but quickly calculable, improvements to the objective function, ultimately speeding convergence.

It is natural to suspect that the alternative $k$-means algorithms can also be adapted for $k$-modes.
Indeed, the $k$-modes algorithm proposed by \citet{Chaturvedi2001} follows the logic of Lloyd's $k$-means algorithm, but to our knowledge no one has adapted Hartigan's or Hartigan and Wong's $k$-means algorithms to $k$-modes.
We propose (Section~\ref{methodology}) an Optimal Transfer Quick Transfer (OTQT) algorithm for $k$-modes analagous to Hartigan and Wong's algorithm for $k$-means.
A simpler variant, the Optimal Transfer (OT) algorithm, is analogous to Hartigan's algorithm.
We prove that the proposed algorithms can achieve better optima by detecting objective-improving moves the other algorithms miss.
We evaluate, in Section~\ref{sec:eval}, the performance of all methods on six test datasets, and more thoroughly in a moderately-sized simulation study.
The paper concludes with some discussion. An online supplement containing additional figures (referenced with the prefix ``S-'') summarizes further experimental results.


\section{Methodology}\label{methodology}
\subsection{Preliminaries}\label{sec:preliminaries}
Let $\mbX = \{\bX_1,\bX_2,\ldots,\bX_n\}$ be a dataset of $n$
observation records, where the $i$th record
$\bX_i=(X_{i1},X_{i2},\ldots,X_{ip})$ has $p$ 
categorical-valued features. The \textit{weighted dissimilarity} between any two
records $\bX_i$ and $\bX_j$ is determined by the number of mismatched features, or specifically,
\begin{equation}
\delta(\bX_i,\bX_j) = \sum_{\ell = 1}^p\varpi_{\ell,X_{i\ell},X_{j\ell}} \1(X_{i\ell} \not\equiv X_{j\ell}),
\label{wt.diss}
\end{equation}
where $X_{i\ell} \not\equiv X_{j\ell}$ implies that $X_{i\ell}$ and
$X_{j\ell}$ are in different categories and $\1(\cdot)$ is an
indicator function taking an event and returning value 1 or 0 according to the truth of the event.
When $\varpi_{\ell,X_{i\ell},X_{j\ell}} \equiv 1$ for all $i,j,\ell$, then
\eqref{wt.diss} reduces to the Hamming distance, giving equal
importance to each observation and category of an attribute.
We use the Hamming distance in this paper, but there has been much discussion about alternative weights or distances~\citep{Ng2007,Bai2011,He2011a,Cao2012,Kim2012,Santos2015,Kim2017a}.

The objective of $k$-modes is to minimize the summed distances of the observations to their assigned cluster modes.
Letting
\begin{equation}
  \Delta(\mbX, \bmu) = \sum_{i=1}^n \delta(\bX_i, \bmu),
\label{mode}
\end{equation}
with $\delta(\cdot,\cdot)$ as in \eqref{wt.diss}, \citet{huang97b} defines a mode of $\mbX$ as the vector $\hat\bmu =(\hat\mu_1,\hat\mu_2,\ldots \hat\mu_p)$ which minimizes \eqref{mode}, \textit{i.e.}~$\hat\bmu = \argmin_{\bmu}\Delta(\mbX,\bmu)$. 
Extending this concept to $K$ clusters, a $k$-modes algorithm must identify the partition $\mbC = \{\mC_1,\mC_2,\ldots,\mC_K\}$ of $\mbX$ and modes $\bmu_1,\bmu_2,\ldots,\bmu_K$, such that the objective function 
\begin{equation}\label{eqn:objfun}
  \mW_K = \sum_{k=1}^K \sum_{i=1}^n  \1(\bX_i \in \mC_k) \delta(\bX_{i},\bmu_{k})
\end{equation}
is minimized. At the minimum value, $\hat\bmu_k$ is the mode of the observations in $\mC_k$.

There are multiple optimization methods that can be developed  to minimize~\eqref{eqn:objfun}.
The $k$-modes algorithm of \citet{huang97b} starts with $K$ initial
modes $\bmu_1,\bmu_2,\ldots,\bmu_K$ and iteratively minimizes
\eqref{eqn:objfun} using the following steps:
\begin{enumerate}
  \item \label{itm:start}
\textbf{Initialize.}
For each $i=1,2,\ldots,n$, allocate $\bX_i$ to the group with closest mode--that is, assign $\bX_i$ to cluster $\mC_{\hat k}$ where $\hat k = \argmin_{k} \delta(\bX_i, \bmu_k)$.
Update $\bmu_{\hat k}$ after each allocation, which by minimizing \eqref{mode}, is
\begin{equation}\label{eqn:mode_update}
	\mu_{\hat k\ell} = \argmax_{c\in\mJ_\ell} \sum_{j=1}^i \1(\bX_j\in\mC_{\hat k}, \bX_{j\ell} = c),
\end{equation}
for the $\ell$th coordinate given the current observation $i$.
We define an arbitrary ordering (using comparators $\prec$ or $\succ$) of the distinct observed characters $\mJ_\ell$ at coordinate $\ell$;
the mode is assigned the low rank category when there is a tie,
guaranteeing its uniqueness. 

  \item \label{itm:reallocation} 
\textbf{Reallocate.}
For each $i=1,2,\ldots,n$, move $\bX_i$ from the current cluster $k$ to another cluster $k'$ if it is closer to $\bmu_{k'}$ than $\bmu_{k}$.
If there is more than one competing cluster at the same minimum distance, we simply assign the observation to the cluster of minimum index.
After each reallocation, update $\bmu_{k'}$ and $\bmu_k$ using \eqref{eqn:mode_update} with upper limit $i=n$ on the sum.

  \item \label{itm:repeat}
\textbf{Repeat.}
Repeat Step~\ref{itm:reallocation} until there are no reassignments during a full cycle through the dataset.
\end{enumerate}

We note a few details about this algorithm.
Affected modes $\bmu_k$ are updated after every allocation, including after each initial assignment in the first iteration.
Especially because of these updates, it can be important 
to shuffle the observation input order before starting the algorithm, since we have observed pathological orderings that fail to achieve the global minimum no matter which algorithm or initialization strategy is used.
The arbitrary category ordering and rules for tie resolution can impact the obtained modes and clusters, but not the optimal value of the objective function.
The same rules will be used in the \hw{} algorithm, and the implications of these choices will be further mentioned in the discussion.

\subsection{A New, More Efficient $k$-Modes Algorithm}\label{sec:otqt}

\citet{hartiganandwong79} make two major improvements on the~\citet{macqueen67} algorithm for $k$-means. 
Specifically, they (a) choose moves to maximally improve the objective function and (b) avoid attempting any move that has no hope to improve the objective function.
They also propose a heuristic \textit{quick transfer stage}, where only transfers between clusters with the closest and the putative next closest means are considered.
We provide a new $k$-modes algorithm in the spirit of \citet{hartiganandwong79}.
The proposed \hw{} algorithm begins with step $1$ of the H97 algorithm and then alternates the \textit{optimal transfer} stage, where each observation is moved to the cluster that most improves the objective function, and the \textit{quick transfer} stage until there are no reassignments during optimal transfer.
Optimal transfers can improve the objective function even when H97 \textit{reallocation} cannot.
\begin{figure}[h]
  \centering
  \vspace{-0em}
  \includegraphics[width=0.45\textwidth]{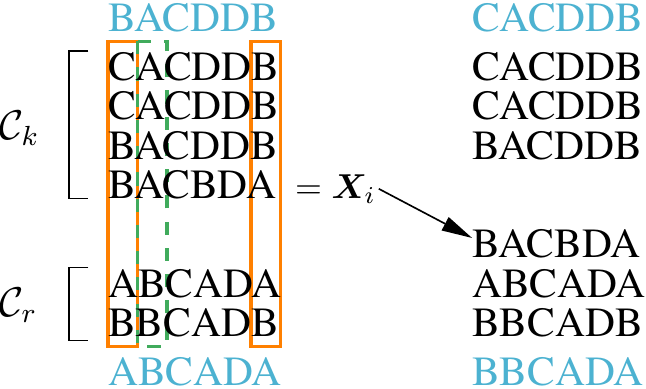}
\caption{A move taken by algorithm \hwo{} but not H97 decreases objective function by $1$.
Observation $\bX_i$ is in cluster $\mC_k$ at Hamming distance $2$ from the mode (blue).
Since it is at Hamming distance $3$ from the mode of cluster $\mC_r$, it will stay in cluster $\mC_k$ under algorithm H97.
However, two sites support the move (boxed in orange) and one site does not (boxed in dashed green), so \hwo{} would make the move, decreasing the cost of these two clusters from $6$ to $5$.}
\label{fig:picture_proof}
  \vspace{-0em}
\end{figure}
In \figref\ref{fig:picture_proof}, H97 detects $\bX_i$ is closer to the mode of $\mC_k$ and leaves it in place, but the optimal transfer stage of \hw{} sees the objective function can be reduced by transferring $\bX_i$ to cluster $\mC_r$.

Before describing the algorithm in detail, we state and prove the cost of moving an observation from cluster $\mC_k$ to $\mC_r$ (Claim~\ref{claim:cost}) needed to detect \textit{optimal transfers}.
An important observation emerging from the proof of Claim~\ref{claim:cost} is that there are two types of sites that determine whether a move will take place:
\begin{enumerate}
	\item\label{itm:volatile} A site $\ell$ that favors the move is one where $X_{i\ell}$ had mode status, or gains it by breaking a tie, in cluster $\mC_r$ \textit{and} did not have mode status, or loses it by breaking a tie, in cluster $\mC_k$ upon departure.

	\item\label{itm:stable} A site $\ell$ that disfavors the move is one where $X_{i\ell}$ did not have mode status, or gains it by creating a tie, in cluster $\mC_r$ \textit{and} had mode status, or loses it by creating a tie, in cluster $\mC_k$ upon departure.

\end{enumerate}
All other sites are irrelevant, and if there are more sites of the
first type than of the second, 
then $\bX_i$ will move from $\mC_k$ to $\mC_r$.
In \figref\ref{fig:picture_proof}, two sites favor and one site disfavors the move that improves the objective function.

For the proof and subsequent algorithm, we define the \textit{minor modes} $\bem_1,\bem_2,\ldots\bem_K$, where
\[
	m_{k\ell} = \argmax_{\substack{c\in\mJ_\ell\\ c\neq \mu_{k\ell}}} \sum_{j=1}^n\1(\bX_j\in\mC_k, \bX_{j\ell} = c)
\]
given the modes $\bmu_k$ for $1\leq k\leq K$.
Further, we maintain up-to-date counts, $n_{k\ell c}$, of category $c$ at coordinate $\ell$ in cluster $k$.
The subsequent discussion may be easier to follow upon noticing that objective function~\eqref{eqn:objfun} can be rewritten in terms of these counts
\[
	\mW_K = \sum_{k=1}^K \sum_{\ell=1}^{p} \sum_{\substack{c\in\mJ_\ell\\ c\neq \mu_{k\ell}}} n_{k\ell c}.
\]
Also from these counts, we quickly compute the $k$th mode, using $\mu_{k\ell} = \argmax_{c\in\mJ_\ell} n_{k\ell c}$, and the $k$th minor mode, using $m_{k\ell} = \argmax_{\substack{c\in\mJ_\ell, c\neq \mu_{k\ell}}} n_{k\ell c}$ for all $\ell$.

\begin{claim}
\label{claim:cost}
The cost of moving observation $\bX_i$ from cluster $\mC_k$ to $\mC_r$ is
\begin{equation*}
  \begin{split}
    \sum_{\ell=1}^p 
    \bigg[&\1\left(n_{r\ell\mu_{r\ell}}
      \geq n_{r\ell X_{i\ell}} + 1\right) 
    - \1\left(X_{i\ell} = \mu_{k\ell}, n_{k\ell\mu_{kl}} = n_{k\ell m_{kl}}\right)
		- \1\left(X_{i\ell} \neq \mu_{k\ell}\right)
              \bigg].
\end{split}
              \end{equation*}
\end{claim}

\begin{proof}
We consider all possible changes at coordinate $\ell$ when moving $\bX_i$ from cluster $k$ to $r$.
In these descriptions, counts $n_{k\ell j}$, modes $\mu_{k\ell}$ and minor modes $m_{k\ell}$ refer to the counts, modes, and minor modes \textit{before} the move.
In addition, we make heavy use of the arbitrary ordering of the elements in $\mJ_\ell$ observed at site $\ell$.
In italics, we highlight the conditions that lead to a change either in the cost of cluster $\mC_r$ or the cost of cluster $\mC_k$.
Compare these italicized conditions to the three events indicated in the equation to complete the proof.

We consider the three possible outcomes at coordinate $\ell$ in cluster $\mC_r$ of moving $\bX_i$ to $\mC_r$: either $X_{i\ell}$ is already the mode, $X_{i\ell}$ is not the mode and will not become the mode, or $X_{i\ell}$ is not the mode, but will become the mode.
Clearly, if $X_{i\ell}=\mu_{r\ell}$ is already the mode, then adding $\bX_i$ only increases the support for the mode and there is no cost of adding $\bX_i$ to cluster $\mC_r$ at site $\ell$.
\textit{If the mode $\mu_{r\ell}\neq X_{i\ell}$ and adding $X_{i\ell}$ does not change the mode $\mu_{r\ell}$, then there will be a cost of $1$ at coordinate $\ell$ for adding $X_{i\ell}$ to cluster $\mC_r$.
This situation occurs if $\mu_{r\ell}$ is observed more often than $X_{i\ell}$ in cluster $\mC_r$, even after adding $\bX_i$, \textit{i.e.}, $n_{r\ell\mu_{r\ell}}>n_{r\ell X_{i\ell}}+1$,
or $n_{r\ell\mu_{r\ell}}=n_{r\ell X_{i\ell}}+1$, but $\mu_{r\ell} \prec X_{i\ell}$.}
Finally, if addition of $\bX_i$ changes the mode at site $\ell$, it can be because $X_{i\ell}=m_{r\ell}$ is a minor mode and $n_{r\ell\mu_{r\ell}} = n_{r\ell m_{r\ell}}$ (with ordering $\mu_{r\ell} \prec X_{i\ell}$), and the cost does not change as there were $n_{r\ell m_{r\ell}}$ mismatches before and there will be an equal $n_{r\ell\mu_{r\ell}}$ mismatches after the change.
\textit{Addition of $\bX_i$ can also change the mode if $n_{r\ell\mu_{r\ell}}=n_{r\ell X_{i\ell}}+1$ and $X_{i\ell} \prec \mu_{r\ell}$, in which case the cost increases by $1$ because there were $n_{r\ell X_{i\ell}}$ mismatches before the change, but there will be $n_{r\ell\mu_{r\ell}}=n_{r\ell X_{i\ell}}+1$ mismatches after the change.}

Meanwhile, there are three possible outcomes of moving $\bX_i$ out of cluster $\mC_k$: either $X_{i\ell}$ is not the mode, it is the mode and will stay the mode, or it is the mode and the mode will change.
\textit{If $X_{i\ell}\neq\mu_{k\ell}$ is not the mode, it cannot become the mode upon $\bX_i$'s departure, and the cost will reduce by $1$ when $\bX_i$ departs.}
If $X_{i\ell}=\mu_{k\ell}$ is the mode and the mode does not change, it is because $n_{k\ell\mu_{k\ell}} > n_{k\ell m_{k\ell}}+1$ or
	$n_{k\ell\mu_{k\ell}} = n_{k\ell m_{k\ell}}+1$ and $\mu_{i\ell} \prec m_{k\ell}$.
In both cases, the cost does not change.
Finally, there are two ways for $X_{i\ell} = \bmu_{k\ell}$ to lose mode status upon departure of $\bX_i$.
If $n_{k\ell\mu_{k\ell}} = n_{k\ell m_{k\ell}} + 1$ and $m_{k\ell} \prec \mu_{k\ell}$, then $m_{k\ell}$ will become the mode, but the new cost will be $n_{k\ell\mu_{k\ell}} - 1 = n_{k\ell m_{k\ell}}$, the old cost.
\textit{Otherwise if there is a tie $n_{k\ell \mu_{k\ell}}=n_{k\ell m_{k\ell}}$ for the mode and $X_{i\ell} \prec m_{k\ell}$, then $\bX_i$'s departure will make $m_{k\ell}$ the mode.
The new cost will be $n_{k\ell\mu_{k\ell}} - 1 = n_{k\ell m_{k\ell}} - 1$, $1$ less than the previous cost.}
\end{proof}
We now provide the details of the proposed \hw{} algorithm.
Let the ``live set'' of clusters be those that have experienced a change in membership during the last cycle through the observations.
We additionally define the ``cost set'' as the collection of clusters whose membership costs (contribution to objective function~\eqref{eqn:objfun}) need to be recalculated.
Given a set of $k$ modes obtained by some initialization procedure,
the \hw{} algorithm is:
\begin{strip}
\begin{enumerate}
\item \label{itm:hwstart} {\bf Getting started.}
For each observation $\bX_i, i=1,2,\ldots,n$, find the two closest
modes $\bmu_k$ and $\bmu_r$. Assign $\bX_i$ to the cluster $\mbC_k$
with closest mode $\bmu_k$, which is then updated.
If the distances to $\bmu_k$ and $\bmu_r, k<r$, are tied, we assign the
observation to $\bmu_k$. 
Place all clusters in the live and cost sets.
Compute the minor modes. At this time, the modes and minor modes for
each group are correct, but the two closest modes to each observation have not
been updated after the assignment of subsequent observations.
As the algorithm progresses, the closest and second closest mode
information will improve. 
\item \label{itm:optra} {\bf Optimal transfer stage.}
Loop through all observations.  For an observation $\bX_i$ in cluster
$k$: 
	\begin{enumerate}
	\item If $k$ is in the cost set, recompute the cost of $\bX_i$'s membership in cluster $k$ from Claim~\ref{claim:cost} as
          \begin{equation}
\label{eqn:cost_of_membership}
\sum_{\ell=1}^p
 \Bigl[\1\bigl(X_{i\ell} = \mu_{k\ell}, n_{k\ell\mu_{k\ell}} =
 n_{k\ell m_{k\ell}}\bigr)
 \mbox{} + \1\bigl(X_{i\ell} \neq \mu_{k\ell}\bigr)\Bigr].
\end{equation}
	\item \label{itm:transfer}
Now check the cost of moving $\bX_i$, while applying a
branch-and-bound algorithm to improve computational efficiency.
Specifically, we first check the cost of moving $\bX_i$ to the recorded second-closest mode, if it is live.
We also check the cost of moving $\bX_i$ to all other live clusters, aborting the calculation as soon as it exceeds the previous lowest cost.
From Claim~\ref{claim:cost}, the cost of adding $\bX_i$ to  $\mbC_r, r\neq k$, is
\begin{equation}\label{eqn:cost_to_join}
\begin{split}
\sum_{\ell=1}^p
	\1(n_{r\ell \mu_{r\ell}} \geq n_{r\ell X_{i\ell}}+1).
\end{split}
\end{equation}
If the minimum of these costs is less than the cost of membership in
the current  $\mbC_k$ or the costs are equal and the new cluster index
$r < k$, {\bf transfer} $\bX_i$ from the old  $\mbC_k$ to the new $\mbC_r$:
		\begin{itemize}
		\item Put $\mbC_k$ and $\mbC_r$ in both the live and
                  cost sets.  
		\item Assign $\bmu_r$ as the closest mode and $\bmu_k$ as the
                  second closest mode to $\bX_i$.
		\item For each coordinate $\ell$, increase  $n_{r\ell X_{i\ell}}$ and decrease  $n_{k\ell X_{i\ell}}$ by one.
		\item We may need to update the source cluster's mode $\bmu_k$ and minor mode $\bem_k$.  For each coordinate $\ell$:
			\begin{itemize}
			\item If $X_{i\ell} = \mu_{k\ell}$, $n_{k\ell \mu_{k\ell}} = n_{k\ell m_{k\ell}}$ and $m_{k\ell} \prec \mu_{k\ell}$, 
set $\mu_{k\ell}=m_{k\ell}$ and 
$m_{k\ell}$ to the lowest ranked category $c$ with $m_{k\ell} \prec c \preceq X_{i\ell}$ and $n_{k\ell c} = n_{k\ell m_{k\ell}}$.
			\item If $X_{i\ell} = \mu_{k\ell}$ and $n_{k\ell \mu_{k\ell}} = n_{k\ell m_{k\ell}} - 1$, then set $\mu_{k\ell} = m_{k\ell}$.  
			Search through categories to set $m_{k\ell}$.
                        It will be the first category $c \succ m_{k\ell}$
                        with $n_{k\ell c}=n_{k\ell m_{k\ell}}$ or, if
                        the former does not exist, then the first
                        category $c\preceq X_{i\ell}$ with $n_{k\ell
                          c}=n_{k\ell X_{i\ell}}$. 
			\item Otherwise, if $X_{i\ell} = m_{k\ell}$,
                          then search through $\mJ_\ell$ to detect a possible new $m_{k\ell}$.
			\end{itemize}
		\item We may also need to update the target clusters's mode $\bmu_r$ and minor mode $\bem_r$.  For each coordinate $\ell$,
			\begin{itemize}
			\item If $n_{r\ell X_{i\ell}} > n_{r\ell \mu_{r\ell}}$ or $n_{r\ell X_{i\ell}} = n_{r\ell \mu_{r\ell}}$ and $X_{i\ell} \prec \mu_{r\ell}$, then set $m_{r\ell} = \mu_{r\ell}$ and $\mu_{r\ell}=X_{i\ell}$.
			\item Otherwise, if $n_{r\ell X_{i\ell}} > n_{r\ell m_{rj}}$ or $n_{r\ell X_{i\ell}} = n_{r\ell m_{r\ell}}$ and $X_{i\ell} \prec m_{r\ell}$, then set $m_{r\ell} = X_{i\ell}$.
			\end{itemize}
		\end{itemize}
	\item
	Remove all clusters from the cost set and remove all clusters not involved in a transfer from the live set.
	The algorithm terminates if the live set is empty.  
	\end{enumerate}
\item \label{itm:qtran} {\bf Quick transfer stage.} Repeatedly loop
  through all observations.  For any observation $\bX_i$ in cluster
  $k$: 
	\begin{enumerate}
	\item Recompute the cost of $\bX_i$'s current membership by \eqref{eqn:cost_of_membership} if $k$ is in the live set.
	\item
If either the cluster $\mbC_k$ with the closest mode or $\mbC_r$ with the second closest mode are in the live set, compute the cost of joining $\mbC_r$ by \eqref{eqn:cost_to_join}.	
If the cost to join $\mbC_r$ is less than the cost of membership in
$k$, carry out transfer in Step~\ref{itm:transfer}, but do not place $\mbC_k$ or $\mbC_r$ in the cost set.	
	\item If there has been no transfer for the last $n$ observations processed, go to the Optimal Transfer Stage (Step~\ref{itm:optra}).
	\end{enumerate}
      \end{enumerate}

      \end{strip}
        
The \hw{} algorithm just described alternates between optimal transfer and quick transfer stages, but
we also consider algorithm \hwo,
which iterates the optimal transfer stage without passage through the quick transfer stage.

\subsection{Theoretical Performance Guarantees\label{sec:theory}}

We now provide some theoretical understanding of how the \hwo{} algorithm works.
Since the quick transfer stage appears to be heuristic~\citep{hartiganandwong79}, we do not include it in this formal evaluation.
We prove the \hwo{} algorithm to be monotone and guaranteed to terminate in a finite number of iterations. 
We show that the clusters are nonempty and maintain distinct modes at each iteration.
As a result, the algorithm is guaranteed to provide a $K$-mode solution if it is initialized with $K$ distinct modes.
Finally, we show that the set of moves made by H97 is a strict subset of the moves made by \hwo.
Thus, \hwo{} is guaranteed to achieve an equal or better update than H97 during every iteration from the same state.

\begin{claim}
\label{claim:2}
The \hwo{} algorithm 
 initialized with $K\leq n$ distinct observations has the following properties:
\begin{enumerate}
\item\label{itm:decrease} The objective function value $\mW_K$ decreases at every iteration.
\item\label{itm:terminate} The algorithm terminates after finitely many iterations.
\item\label{itm:empty} The partition has no empty clusters at any iteration.
\item\label{itm:equal} The modes remains distinct at every iteration.
\end{enumerate}
\label{claim:properties}
\end{claim}

\begin{proof}
We prove each property in turn.
\begin{enumerate}
\item
The algorithm terminates unless there is an observation $\bX_i$ whose
move from cluster $\mbC_k$ to $\mbC_r$ will reduce the objective
function $\mW_K$ by at least $1$, so Property~\ref{itm:decrease} holds.

\item
The objective function is bounded, that is,  $0\leq\mW_K\leq (n-K)p$, where $n$ is the number of observations and $p$ is the number of coordinates.
Thus, the initial value of the objective function is finite, and the algorithm must terminate in finitely many steps.

\item We distinguish empty clusters introduced during initialization from those formed during later iterations.
Recall that during initialization and subsequent iterations, we process the observations in their index order.

There cannot be an empty cluster after the first iteration if there
are no updates during the first iteration. 
If there are updates, each update occurs \textit{after} an observation is assigned to the cluster.
To get an empty cluster, there must be a cluster $\mbC_k$ whose mode is updated to match that of $\mbC_r$ \textit{before} any observations are added to $\mbC_r$.
Then, even the observation that seeded $\mbC_r$ will get assigned to $\mbC_k$ if $k<r$.
Without loss of generality (WLOG), suppose adding $\bX_i$ to $\mbC_k$ precipitates a change in $\bmu_k$ to match $\bmu_r$.
This is a contradiction because before the addition, there must exist a site $\ell$ in $\bmu_k$ where $\mu_{k\ell}\neq X_{i\ell}$, costing $1$, but $\mu_{r\ell}=X_{i\ell}$ costs nothing.
Apart from such sites, all other sites in $\bmu_k$ and $\bmu_r$ either both match or both mismatch $\bX_i$, and thus do not change the distance.

To form an empty cluster at a later iteration, its last member must move to a new cluster.
Since the cost of membership of this last member is $0$, there can be no lower or equal cost of membership in any other cluster, unless there is another cluster with an identical mode.
Claim~\ref{claim:2}.\ref{itm:equal} disallows the possibility of equal modes.

\item 
The argument in proving~\ref{itm:empty} also guarantees $K$ distinct modes \textit{after the first iteration}, which we will use in the following proof.
We separate the proof below into separate cases for $K=2$ and $K>2$ modes.

If $K=2$, then the modes will remain distinct.
For $\bX_i$ to move between two clusters, there must be at least one site $\ell$ where $X_{i\ell}$ is or becomes the mode in the receiving cluster (without ties), but was not so or loses mode status by breaking a tie in the source cluster (see Claim~\ref{claim:cost}), which contradicts the possibility of equal modes after the move.

For $K>2$, suppose that we obtain two equal modes by moving observation $\bX_i$ from cluster $\mbC_k$ to $\mbC_s$ such that after the move $\bmu_k=\bmu_r$ or $\bmu_s=\bmu_r$ for some $r\notin\{k,s\}$.
At the time of the move, this configuration implies that the minor
mode $\bem_k = \bmu_r$ or $\bem_s = \bmu_r$, and that 
there exists a nonempty subset $\mL$ of
sites where there is a tie or about to be one 
with the mode state in $\mbC_k$ or $\mbC_s$, WLOG $\mbC_k$.
Such sites are said to be in the \textit{cusp state}.
If $\ell\in\mL$, then the mode and minor mode are tied ($n_{k\ell m_{k\ell}} = n_{k\ell\mu_{k\ell}}$ with $\mu_{k\ell} < m_{k\ell}$) or nearly tied ($n_{k\ell m_{k\ell}} = n_{k\ell\mu_{k\ell}} + 1$ with $m_{k\ell} < \mu_{k\ell}$).
For sites $\ell\notin\mL$, the plurality of observations in $\mbC_k$ must match $\bmu_{s\ell}$.

To create the cusp state at $\ell\in\mL$ in cluster $k$ during initialization, we must add an observation $\bX_{j}$ to cluster $k$ although $X_{j\ell} \neq \mu_{k\ell}$.
If $\bX_{j}$ is to go to cluster $k$ instead of any other cluster $h$, it does so either because of a \textbf{tie resolution} or a \textbf{driving difference}.
A \textbf{tie resolution} occurs if $\delta(\bX_{j},\bmu_k) \leq
\delta(\bX_{j},\bmu_h)$ for all $h$ and equality is achieved for some $h$, but $k \leq h$ for all such $h$.
In particular, if $\delta(\bX_{j},\bmu_k) = \delta(\bX_{j},\bmu_r)$ for ultimately tied cluster $r$, either $X_{j\ell} \neq \mu_{r\ell} \neq \mu_{k\ell}$ or $X_{j\ell} = \mu_{r\ell} \neq \mu_{k\ell}$.
In the first case, $\ell$ may end up in the cusp state, but then $m_{k\ell} = \mu_{r\ell}$ cannot be true, as required.
In the second case, there must exist another site $\ell_1\neq\ell$, where $X_{j\ell_1} = \mu_{k\ell_1} \neq \mu_{r\ell_1}$.
The second case also arises if $\bX_j$ is assigned to $k$ instead of $r$ because of a \textbf{driving difference} making $\delta(\bX_{j},\bmu_k) < \delta(\bX_{j},\bmu_r)$.
Since $X_{j\ell} \neq \mu_{k\ell}$, there must be sites elsewhere, in particular some $\ell_1$, where $X_{{j}\ell_1} = \mu_{k\ell_1} \neq \mu_{r\ell_1}$.
If $\ell_1\in\mL$, then it cannot yet be in the cusp state because it was already the mode in cluster $k$ prior to the move.
Otherwise, if $\ell_1\notin\mL$, then $\mu_{k\ell_1} \neq \mu_{r\ell_1}$.
Thus, after addition of $\bX_{j}$, clusters $k$ and $r$ are not in the desired configuration.
Subsequent additions required to achieve the desired configuration follow a similar pattern.

To create the desired configuration after initialization, we must add to \textit{or retain in} cluster $k$ an observation $\bX_j$ even though $X_{j\ell} \neq \mu_{k\ell}$ \textit{after} the move.
Whether we add or retain $\bX_j$ in cluster $k$ is mostly semantic: retaining $\bX_j$ is equivalent to removing it and adding it back.
The \textbf{driving sites} that determine whether we add $\bX_j$ to cluster $k$ as opposed to any other cluster $h$ are those that match the mode in cluster $k$ or become the mode in cluster $k$ \textit{by breaking a tie}.
Any site $\ell\in\mL$ cannot be a driving site for cluster $k$, for then $\mu_{k\ell} = X_{j\ell}$ after the move.
Therefore, there is no driving site or there is another driving site $\ell_1\notin\mL$.
If there is no driving site, then $\bX_j$ must already be in cluster $k$, but the final configuration has not been achieved (see below).
Otherwise, site $\ell$ may end in the desired configuration but $\bmu_{k\ell_1} \neq \bmu_{r\ell_1}$.

We have been able to put cluster $k$ in the cusp state, but
$\mu_{k\ell_1} \neq \mu_{r\ell_1}$ for some $\ell_1\notin\mL$ (Case A)
or $m_{k\ell} \neq \mu_{r\ell}$ for some $\ell\in\mL$ (Case B). We
address each case separately.

\textbf{Case A:}.  To achieve the final state, we must add or retain $\bX_j$ in cluster $r$ even though $X_{j\ell_1} = \mu_{k\ell_1}$.
Therefore, $\ell_1$ may be a driving site for moving $\bX_j$ to
cluster $k$, but not to $r$.
Whether a driving site or not, there must exist another driving site $\ell_2$ to drive or retain $\bX_j$ in cluster $r$.
Such sites contradict the final configuration unless $\bX_j$ is in cluster $k$ and $X_{j\ell_2}$ is not the mode in cluster $k$ or loses the mode status by breaking a tie, but both contradict the presumption that cluster $k$ is already in the cusp state.

\textbf{Case B:}. If cluster $k$ is in the cusp state, but $\mu_{r\ell} \neq m_{k\ell}$, then 
cluster $r$ must attain or be already in the same cusp state as cluster $k$ at site $\ell$.
However, cluster $k$ achieved the cusp state at site $\ell$ through a tie resolution in order to be in Case B.
It is therefore impossible for cluster $r$ to achieve the same cusp state without additional driving sites.

In all cases, it is impossible to achieve the configurations that are
one step away from achieving equal modes. Thus, it is impossible to
achieve equal modes.

\end{enumerate}
\end{proof}

Our next claim demonstrates that given a starting configuration, there
exist scenarios where H97 (or Chaturvedi's algorithm) will stop, but \hwo{} 
will continue and improve the value of the objective function.
For these initializations, \hwo{} will obtain a lower value of the objective function.

\begin{claim}\label{claim:3}
There exist moves $\bX_i$ from $\mbC_k$ to $\mbC_r$ that algorithm
\hwo{} will take, but H97 will not.
\end{claim}

\begin{proof}
For observation $\bX_i$ to move from $\mbC_k$ to $\mbC_r$, there must exist at least one site $\ell$ where $X_{i\ell}$ is or breaks a tie to become the mode in cluster $\mbC_r$ \textit{and} $X_{i\ell}$ was not or broke a tie to lose the mode status in cluster $\mbC_k$.
If at least one of these sites was already the mode in $\mbC_r$ and was already not the mode in $\mbC_k$, then $\delta(\bX_i,\bmu_k) > \delta(\bX_i,\bmu_r)$ and H97 will take the move.
However, if all these sites break a tie to become the mode in $\mbC_r$ \textit{or} break a tie to lose mode status in $\mbC_k$, then $\delta(\bX_i, \bmu_k) \leq \delta(\bX_i, \bmu_r)$ and H97 will not take the move.
\end{proof}

\subsection{Initialization}

The $k$-modes algorithm is iterative and starts with $K$ initial modes 
that have a large impact on the achieved minimum, which is a local minimum, in the case of \hw{} or \hwo{}, and may not even be a local minimum in the case of H97.
Our goal is to reach the global minimum of the $k$-modes objective function~\eqref{eqn:objfun}, which can only be achieved with these algorithms through repeated random initialization.
We simply initialize with modes obtained by randomly selecting $K$ distinct observations from the dataset.
This initialization method is also employed in the \texttt{R} package \texttt{klaR} that implements existing $k$-modes algorithms.


\section{Performance Evaluations}\label{sec:eval}

To evaluate the proposed algorithms, we analyze six real and many simulated datasets.
Before delivering the results in Section~\ref{experiments}, we describe the datasets and our methods for comparison in Section~\ref{sec:datasets}.
\subsection{Experimental Framework} \label{sec:datasets}
\subsubsection{Real-World Data}\label{sec:real}
\begin{table*}
\caption{Real-world datasets.
Here, $n$ is number of observations, $p$ is the number of non-constant coordinates, $p'$ is the number of dimensions explaining 99.9\% of the dataset variation, $K_t$ is the reported true number of clusters, $c=\sum_j (\lvert\mJ_j\rvert - 1)$, where $\lvert\mJ_j\rvert$ is the size of the set of categories observed at coordinate $1\leq j\leq p$, and ARI is the mean Adjusted Rand Index~\protect\citep{hubertandarabie85} for the best solutions at the indicated number of clusters.
The maximum ARI is bolded for each dataset, but if the maximum did not occur for $K\in\{K_t-1,K_t,K_t+1\}$, we report the maximum ARI and the maximizing $K$ in parentheses under column max.}
\label{tab:real}
\centering
  \begin{tabular}{cc|ccccc|cccc}\hline
&       &       \multicolumn{5}{c|}{{\bf Data Dimensions}}   &
                                                                  \multicolumn{4}{c}{{\bf ARI}} \\ 
{\bf Dataset}        & {\bf Abbr.}      & $\boldsymbol{n}$      & $\boldsymbol{p}$      & $\boldsymbol{p'}$     & $\boldsymbol{K_t}$      & $\boldsymbol{c}$      & $\boldsymbol{K_t-1}$    & $\boldsymbol{K_t}$      & $\boldsymbol{K_t+1}$	& max  \\
\hline
cancer		& c     & 699           & 8     & 7.8  & 2     & 72    & 0     & {\bf 0.67}	& 0.39		& -- \\ 
mushroom	& m     & 8,124         & 20    & 15.0 & 2     & 92    & 0     & {\bf 0.61}	& 0.21		& -- \\ 
senators	& se    & 100           & 542   & 49.9 & 3     & 1011  & 0.04	& 0.45 		& {\bf 0.49}	& -- \\ 
soybean		& so    & 47            & 21    & 15.5 & 4     & 40    & 0.65	& {\bf 0.95}	& 0.79		& -- \\ 
splice		& sp    & 3,175         & 60    & 59.3 & 3     & 182   & 0.01	& 0.02 		& 0.03		& {\bf 0.05} (7) \\ 
zoo		& z     & 101           & 16    & 14.5 & 7     & 26    & 0.63	& 0.66		& 0.63		& {\bf 0.83} (5) \\ 
\hline
\end{tabular}
\end{table*}

We use six real-world categorical datasets to test the new algorithms (Table~\ref{tab:real}).
Our analysis of each dataset is not meant to be ideal nor even correct, but evaluating a variety of real datasets tests the algorithms under multiple, potentially complicated conditions not easy to mimic in simulation.
Five of the datasets are from the UCI Machine Learning Repository~\citep{Lichman:2013}.
The breast cancer (c) dataset contains missing values in one coordinate, which we drop prior to our analysis\,\footnote{This breast cancer domain was obtained from the University Medical Centre, Institute of Oncology, Ljubljana, Yugoslavia. Thanks go to M.~Zwitter and M.~Soklic for providing the data.}.
Several of the coordinates are better considered as ordinal features, but the dataset is classified as categorical in the repository and we treat it as such.
The mushroom (m) dataset contains missing values in one coordinate and no variation at a second: we drop both coordinates prior to analysis.
The soybean (so) dataset, which is the \textit{small} soybean dataset from the repository, contains $14$ coordinates with no variation that we drop prior to our analysis.
The zoo (z) dataset includes one numeric coordinate, number of legs, that we treat as categorical.
The senators dataset (se) \citep{banerjeeetal08} is the voting patterns of 100 senators in the 109th US Congress from January 3, 2005 to January 3, 2007. 
This dataset has 542 coordinates, each of which has three categories (voting for, against, or no vote recorded).
There were nominally three groups of senators (44 Democrats, 1 Independent and 55 Republicans), so $K=3$, however, the independent senator caucused with the Democratic senators during this Congress.
\begin{figure}[h]
\centering
  \includegraphics[width=.5\textwidth]{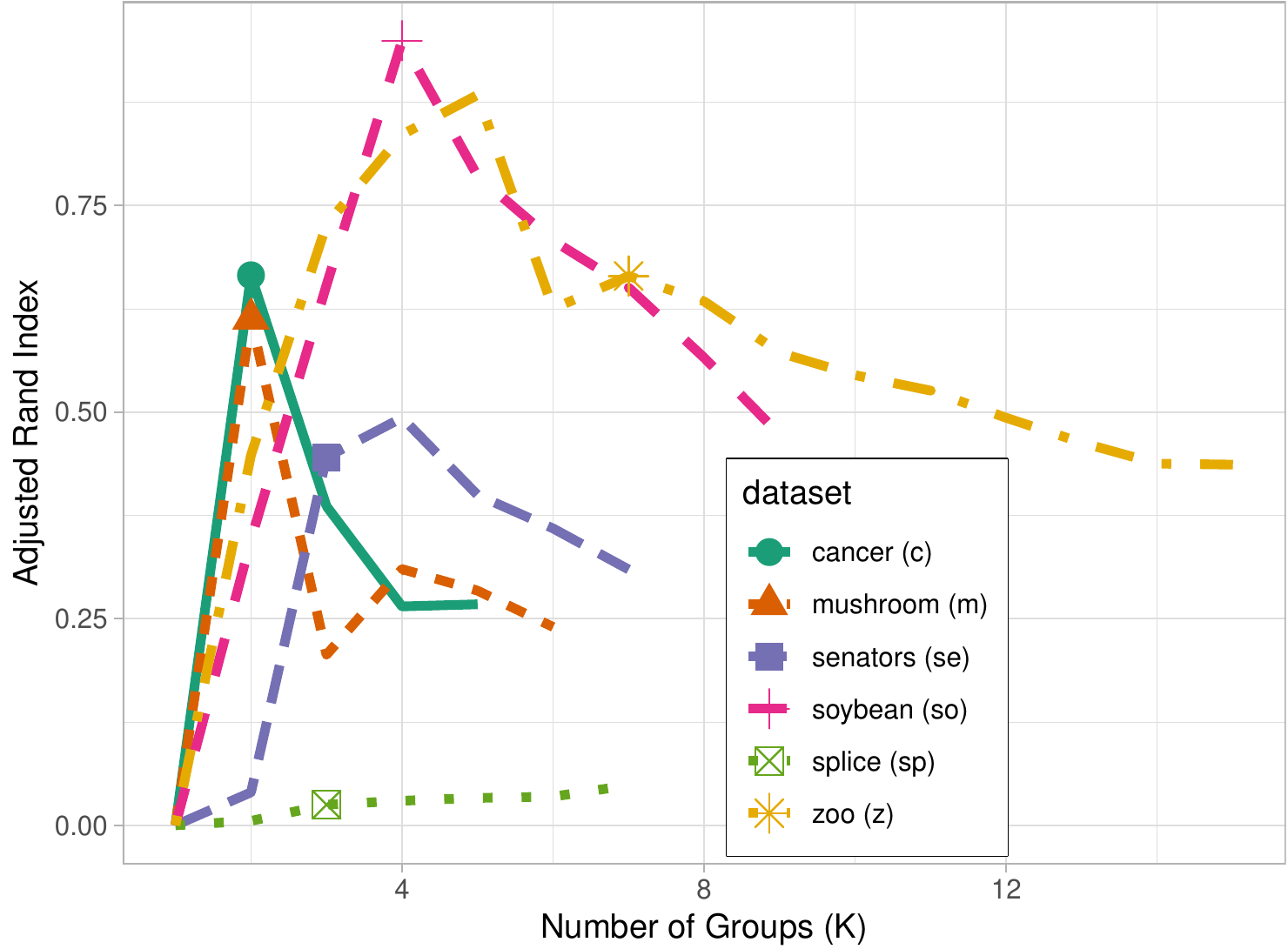}
\caption{
The mean achieved ARI at each tested number of clusters $K$.
The single point along each trajectory indicates the reported true $K_t$.
\label{fig:real}}
\end{figure}

To characterize the strength of the clustering signal in these datasets under the $k$-modes objective function, we compute the average Adjusted Rand Index~\protect\citep{hubertandarabie85} (ARI in Table~\ref{tab:real}) for all solutions that achieved the global minimum for each value of $K$ in a range of choices (see Fig.~\ref{fig:real}).
Since we do not know the true global minimum, we assume the minimum achieved criterion across all algorithms and random initializations in this study is the global minimum.
Except in the splice dataset for $K>2$, the observed minimum is achieved several times.
The ARI for solutions achieving the global minimum may vary because the optimization function is discrete and there is randomness in the clustering of observations with tied distances to more than one mode.
Four datasets achieved a maximal ARI at or near the reported true number of clusters $K_t$ (Fig.~\ref{fig:real}).
The splice dataset achieved a maximum ARI of $0.05$ at $K=7$, well above the reported value of $K_t=3$.
The low ARI values indicate little to no structure in the splice data, at least no structure consistent with the stated labels or the $k$-modes model.
The zoo data achieved a maximum ARI of $0.88$ at $K=5$, a strong indication of clustering, but below the reported value of $K_t=7$.
Thus there may be insufficient information to distinguish all seven reported clusters, but the labels appear to be consistent with the detectable structure in the zoo data.

\subsubsection{Simulation}\label{sec:simulation}
We undertook a simulation study to explore the behavior of the proposed \hw{} and \hwo{} algorithms and the existing H97 algorithm.
We simulated $20$ replicate samples of size $n=1000$ with $4$ categories possible at each of $p\in\{5,10\}$ coordinates, $K\in\{2,5\}$ clusters, and four levels of clustering difficulty (Figure~\ref{fig:simulation}\,\subref{tab:simulation_conditions}).

\begin{figure}
  \centering
  \mbox{
  \subfloat[Structure of simulation data\label{fig:simulation_structure}]{
\raisebox{-.35\height}{\resizebox{2.1in}{2.1in}{%
\tikzstyle{level 1}=[sibling angle=72, level distance=30mm]
\tikzstyle{level 2}=[sibling angle=30, level distance=13mm]
\tikz [grow cyclic, shape=circle,very thick,cap=round]
\node [fill] (root) {}
        child [color=red]
        { node [fill] (red) {}
                child [color=red!10!white]
                { node [fill] {} edge from parent node [left=-3pt,
                  black, pos=0.6] {\Large $t$} }
                child [color=red!\B!\C] foreach \B/\C in {30/white,50/white,70/white,90/white,90/black,70/black,50/black,30/black}
                { node [fill] {} } edge from parent node
                [left,black,pos=0.43] {\Large $t_0$}
        }
        child [color=\A] foreach \A in {green,blue,orange,purple}
        { node [fill] (\A) {} child [color=\A!\B!\C] foreach \B/\C in {10/white,30/white,50/white,70/white,90/white,90/black,70/black,50/black,30/black}
                { node [fill] {} }
        };

}}
}}
\mbox{
\subfloat[Simulation conditions\label{tab:simulation_conditions}]{
\centering
\setlength{\tabcolsep}{3pt}
\begin{tabular}{cccc|cccc}\hline
	&	&	& {\bf Change}		& \multicolumn{4}{c}{{\bf ARI ($\boldsymbol{p,K}$)}} \\
$\boldsymbol{p}$	& $\boldsymbol{K}$	& $\boldsymbol{t}$	& {\bf Prob.}		&  $\boldsymbol{5,2}$	& $\boldsymbol{5,5}$		& $\boldsymbol{10,2}$	& $\boldsymbol{10,5}$ \\
\hline
$\{5,10\}$	& $\{2,5\}$	& $0.5$	& $0.12$	& $0.996$	& $0.995$	& $0.971$	& $0.925$ \\
$\{5,10\}$	& $\{2,5\}$	& $1.0$	& $0.21$	& $0.976$	& $0.959$	& $0.875$	& $0.785$ \\
$\{5,10\}$	& $\{2,5\}$	& $1.2$	& $0.25$	& $0.979$	& $0.934$	& $0.855$	& $0.662$ \\
$\{5,10\}$	& $\{2,5\}$	& $2.0$	& $0.36$	& $0.834$	& $0.777$	& $0.664$ 	& $0.480$\\
\hline
\end{tabular}
}
}
\caption{Simulations.
\protect\subref{fig:simulation_structure} Nested simulation structure
models $K$ modes (here, $5$) as independent realizations of a CTMC
initialized with the center (black) observation and evolved for time $t_0$.
The observations are then simulated as independent realizations of a second CTMC initialized from each mode and evolved for time $t$.
Better separation of clusters, and easier data to cluster, are  produced as $t_0/t$ increases.
\protect\subref{tab:simulation_conditions} $p$ is the number of coordinates, $K$ is the true number of clusters, and $t$ is the aforementioned indication of clustering difficulty ($t_0$ is held constant).
For each $t$, there is an induced Change Prob., the expected proportion of changed coordinates in an observation, relative to the mode.
We also report the median, among the $20$ replicates, of ARI at the true $K$ (see Methods).
\label{fig:simulation}}
\end{figure}

We allow unequal-sized clusters by simulating the cluster proportions as $\bpi\sim\text{Dirichlet}(1,\allowbreak1, \ldots,\allowbreak 1)$.
Then, the cluster sizes $(\vert\mC_1\vert,\vert\mC_2\vert,\ldots,\vert\mC_K\vert)$ follow a Multinomial$(1000, \bpi)$ distribution.
To guarantee $K$ clusters, we discard any simulation with $\vert\mC_k\vert=0$ for any $1\leq k\leq K$.
To simulate the observations, we assume a nested continuous time Markov chain (CTMC) model.
This model is flexible, but we use a particularly simple formulation, where the rate of change between categories is equal.
In this case, choosing realization times is equivalent to choosing a probability of change, and conditional on change, all substitutions are equally likely.
In the inner level, we simulate $K$ modes from an ``ancestor'' observation.
The ancestor is generated with independent coordinates and uniform category probabilities.
Then, each coordinate of each mode is sampled by initializing an independent realization of the CTMC from the corresponding ancestor state.
In our simulations, the inner CTMC results in a $70\%$ chance of change at each coordinate.
To guarantee $K$ distinct modes, we discard any simulation with two or more identical modes.
Finally, we simulate the $\vert\mC_k\vert$ independent observations within cluster $k$ by applying independent CTMC realizations to each coordinate for $t$ time units.
In our simulations, we explore four values of $t\in\{0.5,1,1.2,2\}$, resulting in probabilities of change that vary from $12\%$ to $36\%$ (Figure~\ref{fig:simulation}\,\subref{tab:simulation_conditions}, ``Change Prob.'').
These simulation conditions produce data with varying difficulty of clustering.
To illustrate, we report the median of mean ARIs~\citep{hubertandarabie85}.
Specifically, for each simulated dataset, we compute the mean ARI observed for solutions achieving the smallest observed criterion~\eqref{eqn:objfun} at the true $K$.
The median across the $20$ simulated datasets is reported in Figure~\ref{fig:simulation}\,\subref{tab:simulation_conditions}.
Clustering difficulty increases with $t$, $K$, and $p$.

\subsubsection{Analysis Methodology}\label{sec:analysis}
We analyze both real and simulated datasets by initializing each of the three algorithms (H97, \hw, and \hwo) $\alpha nKp'$ times. 
Specifically, 
all three algorithms are initialized with the same $\alpha nKp'$ sets of initial modes.
This experimental structure creates dependence between the algorithm results, which is accommodated throughout the analyses that follow.
$K$ ranges in $\{1,2,\ldots,2K_t+1\}$ for the real datasets (except mushroom, where the upper limit is $2K_t+2$, because of a local peak in the ARI at $K=4$, \figref\ref{fig:real}) and $\{1,2,\ldots,2K_t-1\}$ for the simulated datasets with $K_t=5$ and $\{1,2,\ldots,2K_t\}$ for the simulated datasets with $K_t=2$, where $K_t$ is the reported or true number of clusters.
The multiplier $\alpha$ and thus the number of runs (independent initializations) is chosen to be the smallest integer multiple such that $\alpha np'\max\{K\}$ exceeds 500,000.
For the real data, $n$ and $p'$ are the data dimensions as reported in Table~\ref{tab:real}.
For simulation, $n=$ 1,000, $p'\in\{5,10\}$, and $K_t\in\{2,5\}$ as reported in Figure~\ref{fig:simulation}\,\subref{tab:simulation_conditions}.
Since $p'=p$ for the simulated data, we may interchangeably use $p$ and $p'$ when discussing simulation data.

We compare the algorithms 
 in terms of both accuracy per initialization and time to reach a target value of the objective function~\eqref{eqn:objfun}.
Ideally, the target value is the global minimum, but this value is unknown, even for the simulated data.
For the real data, we assume the minimum achieved across all algorithms 
 is the global minimum.
This target is achieved frequently enough to perform statistical tests, except in the splice dataset.
In this case and for all simulation datasets, we consider the target met if the minimum is at or below the 5th percentile of optima achieved across all 
 algorithms.
Henceforth, we refer to this value of the objective function as the ``target''.

To compare the accuracies of pairs of algorithms, we count the number of initializations that achieve the target.
Given 
 a pair of algorithms to compare, we count the number of times the first algorithm achieves the target and the second does not ($n_{10}$) and vice versa ($n_{01}$).
If there is no difference between algorithms, then conditional on $N_{10}+N_{01}=n_{10}+n_{01}$, $N_{10} \sim \text{Bin}(n_{10}+n_{01}, 0.5)$.
To avoid the situation where $n_{10}+n_{01}=0$, we approximate this distribution as $\text{Bin}(n_{10}+n_{01}+2, 0.5)$ and assess the left tail probability as $\Pr(N_{10} \leq n_{10} + 1)$.
Discrepancy between the true distribution and this approximation will be highest for low counts and consequently, yield less significant test results.
To retain the precision of extreme tail probabilities both near $0$ and near $1$, when $n_{10}>n_{01}$, we instead compute $\Pr(N_{01}\leq n_{01}) = 1 - \Pr(N_{10}\leq n_{10}+1)$.
We use Holm's method~\citep{Holm1979} to correct the probabilities for multiple testing, then cap the resulting values at $0.5$ (Holm's method caps them at $1$).
Finally, we negate the values with $n_{10} > n_{01}$ and apply a probit-like transformation $-\text{sign}(x)\Phi^{-1}(\vert x\vert)$ to map the values to the real line, such that 
data supporting the superiority of the first algorithm approach $\infty$, while 
data supporting the superiority of the second algorithm approach $-\infty$.
To quantify the effect size, we compute confidence intervals for rate ratio $r = \frac{n_{10}+1}{n_{01}+1}$ using 
\begin{equation*}
  \begin{split}
    \bigg(r\exp\bigg[-q_{0.975}&\sqrt{\frac{1}{n_{10}+1} +
      \frac{1}{n_{01}+1}}\bigg],
    r\exp\bigg[q_{0.975}\sqrt{\frac{1}{n_{10}+1} +
        \frac{1}{n_{01}+1}}\bigg]\bigg),
    \end{split}
\end{equation*}
where $q_{0.975}$ is the $(1-\alpha/2)$ quantile of the standard normal distribution with $\alpha=0.05$.


To compare the speed of the methods, we record the real time in seconds, also known as ``wait times'', between initializations that strike the target.
To compare algorithms, if the first algorithm strikes the target with a wait time of $T_1$ and the second algorithm strikes the target with a wait time of $T_2$, then we perform a Wald's test of equal mean wait times, $\E[T_1] = \E[T_2]$, using statistic $({\overline T_1 - \overline T_2})/{\sqrt{\text{Var}(\overline T_1 - \overline T_2)}}$,
where $\text{Var}(\overline T_1 - \overline T_2) = \text{Var}(T_1)/n_1 + \text{Var}(T_2)/n_2 - \frac{2}{n_1n_2}\sum_{i=1}^{n_1}\sum_{j=1}^{n_2}\text{Cov}(T_{1i}, T_{2j})$, $T_{1i}$ is the $i$th of $n_1$ observed wait times for method $1$ and $T_{2j}$ the $j$th of $n_2$ observed wait times for method $2$.
The variances are estimated from the sample variance of the observed wait times.
For the covariance, we assume that pairs of initializations are
independent and identically distributed, and therefore the covariance
between any two wait times is determined by the number of
initializations they share in common, so 
\begin{equation*}
\begin{split}
  \text{Cov}(T_{1i}, T_{2j}) = c_{12}&\1\left\{\min\{r_{i1},r_{j2}\} >
  \max\{l_{i1},l_{j2}\}\right\}
  \left(\min\{r_{i1},r_{j2}\} - \max\{l_{i1},l_{j2}\}\right),
\end{split}
\end{equation*}
where $c_{12}$ is the estimated covariance of the paired times to complete one initialization, estimated from the $\alpha np'K$ initializations, 
$l_{il}$ and $l_{jl}$ are the indices of the previous initialization to hit the target (or $0$ if none), and $r_{il}$, $r_{jl}$ are the indices of the $i$th and $j$th strike on the target for $l\in\{1,2\}$ indicating the method.
Confidence intervals for time to target ratio ${\overline T_1}/{\overline T_2}$ are computed via Fieller's method~\citep{Fieller1954}.



\subsection{Results\label{experiments}}
We now compare the $k$-modes algorithms on real and simulated datasets.
It is not our purpose to advocate for the $k$-modes objective function. 
The comparison of objective functions and their relative performance on these and other datasets is for others to evaluate~\citep{Anderlucci2014}.
We have proven that the \hwo\ algorithm is more accurate per iteration, but we must empirically show whether it is more accurate per initialization.
It is possible that finding better moves in early iterations can trap the \hwo\ algorithm in local minima.
Since the \hw\ and \hwo\ algorithms are more expensive per iteration, we also empirically assess speed to the optimum.
The measured speeds are dependent on the particular implementation of
the algorithm, the architecture and the compiler, but provide some
indication of the tradeoff between the computational simplicity of
H97~\citep{huang97b} vs.~the guarantees of \hw. We will show that \hw{} and \hwo{} are more accurate per initialization and almost always faster to the target than the H97 algorithm.

\begin{table}
\caption{
Real data at the true $K_t$.
Average number of initializations and time, in seconds, between initializations achieving the global minimum for algorithms H97, \hw, and \hwo{} (minimum bolded).
$\dag$\hw{} significantly different from H97.
$\ddag$\hwo{} significantly different from \hw.
$\S$Splice data did not repeatedly reach the minimum observed objective value, so all values are computed for initializations achieving minima at or below the 5th quantile of all achieved minima.
}
\label{tab:real_results}  
\centering
\setlength{\tabcolsep}{1.5pt}
\begin{tabular}{@{}c| S[detect-weight, mode=text, table-format=1.1] S[detect-weight, mode=text, table-format=1.1] S[detect-weight, mode=text, table-format=1.1] c| S[detect-weight, mode=text, table-format=2.1] S[detect-weight, mode=text, table-format=1.1] S[detect-weight, mode=text, table-format=1.1] c}
\toprule
			& \multicolumn{4}{c|}{\textbf{Avg.~Inits.~to Minimum}}
			& \multicolumn{4}{c}{\textbf{Avg.~Time to Minimum~(s)}}
\\
\textbf{Data}	
			& \multicolumn{1}{c}{\scriptsize\textbf{H97}}	& \multicolumn{1}{c}{\scriptsize\textbf{\hw}}	& \multicolumn{1}{c}{\scriptsize\textbf{\hwo}}	& \multicolumn{1}{c|}{\scriptsize\textbf{Scale}}
			& \multicolumn{1}{c}{\scriptsize\textbf{H97}}	& \multicolumn{1}{c}{\scriptsize\textbf{\hw}}	& \multicolumn{1}{c}{\scriptsize\textbf{\hwo}}	& \scriptsize\textbf{Scale}
\\
\midrule

c	
	& \B 2.1	& 2.1$\dag$	& 2.1	& $10^0$
	& 3.0	& 2.8$\dag$	& \B 2.7$\ddag$	& $10^{-4}$	
\\

m	
	& 2.1	& \B 2.1$\dag$	& \B 2.1	& $10^0$
	& 12	& 7.7$\dag$	& \B 7.5$\ddag$	& $10^{-2}$	
\\

se	
	& 1.1	& \B 1.1$\dag$	& 1.1	& $10^0$	
	& 1.4	& 1.0$\dag$	& \B 1.0$\ddag$	& $10^{-3}$	
\\

so
	& 2.0	& \B 1.7$\dag$	& 1.7$\ddag$	& $10^0$
	& 1.2	& 1.1$\dag$	& \B 1.0$\ddag$	& $10^{-4}$	
\\

sp$\S$
	& 1.9	& 1.7$\dag$	& \B 1.7$\ddag$	& $10^1$
	& 2.2	& 3.0$\dag$	& \B 2.1$\ddag$	& $10^{-1}$	
\\

z
	& 1.2	& 0.9$\dag$	& \B 0.9	& $10^2$
	& 16	& 8.7$\dag$	& \B 8.2	& $10^{-2}$	
\\
\bottomrule
\end{tabular}
\end{table}

\subsubsection{Real-World Data}
\begin{figure*}
\centering
\vspace{-1em}
\mbox{
  \subfloat[Accuracy\label{fig:real:nini:h97hw}]{\includegraphics[width=0.485\textwidth]{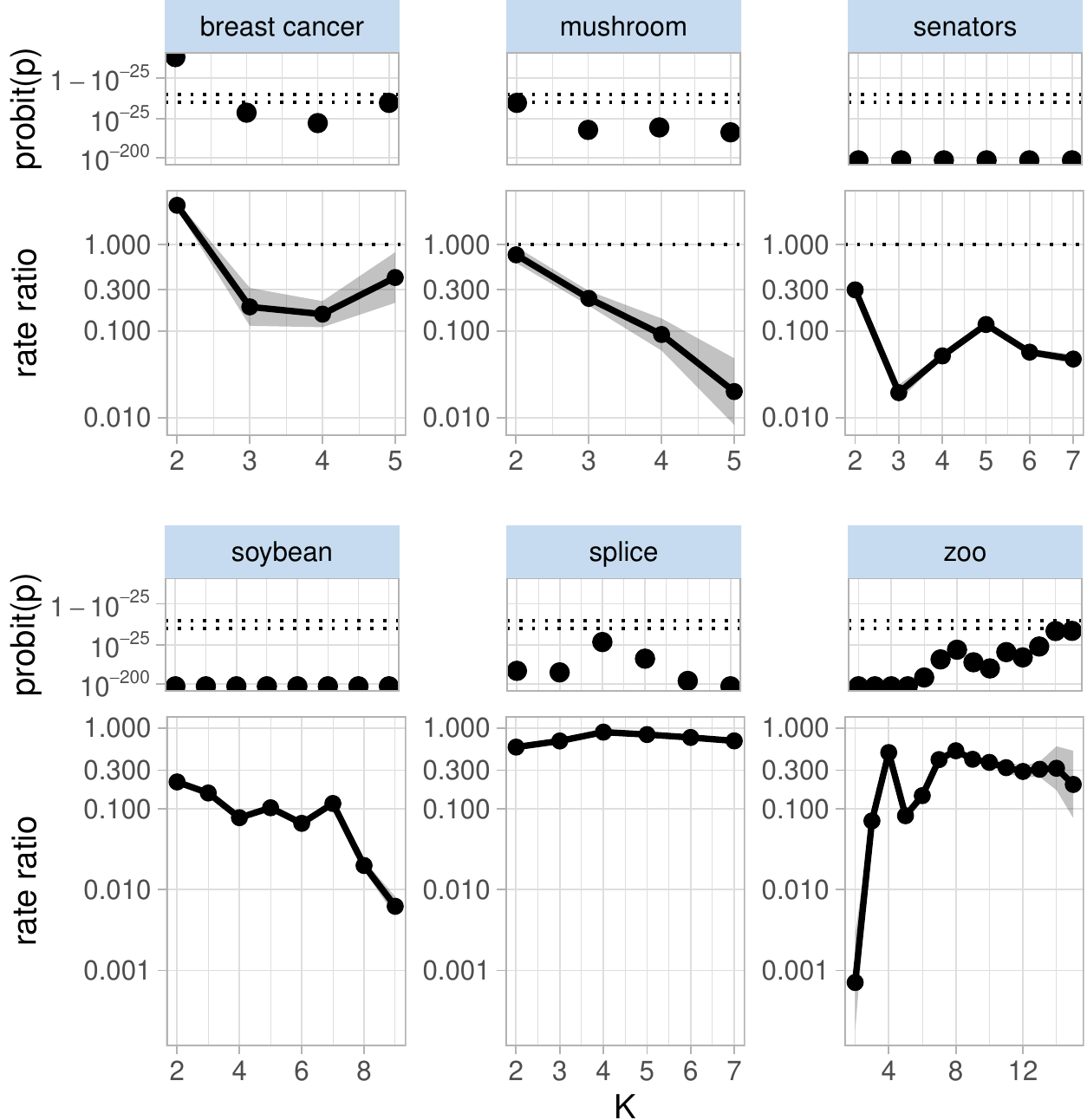}}
\hspace{1em}
  \subfloat[Timing\label{fig:real:time:h97hw}]{\includegraphics[width=0.485\textwidth]{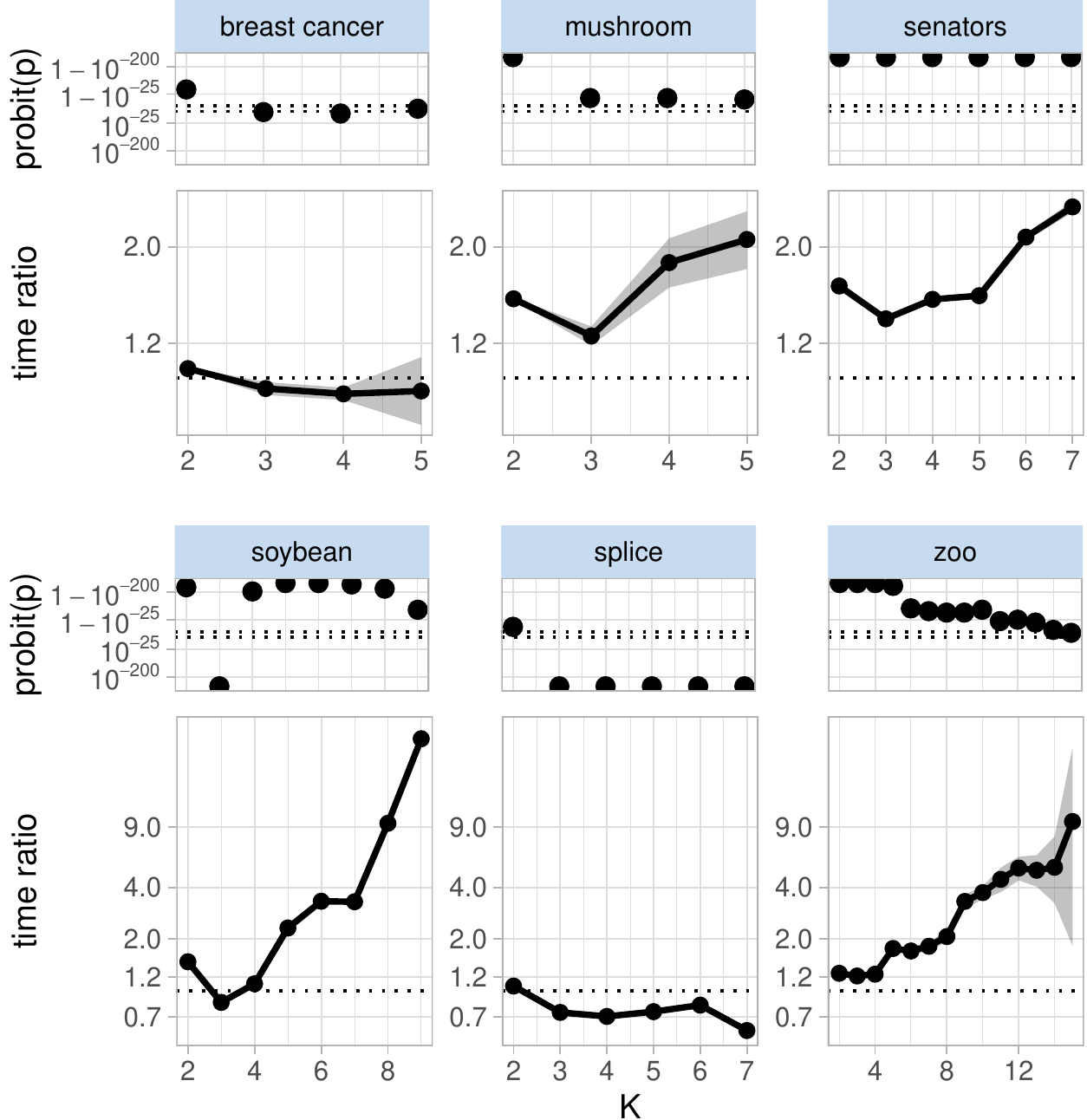}}}
\caption{H97 vs.~\hw\ on real data.
In 12 pairs of plots, top show probit-transformed, adjusted left tail probabilities for tests of equal \protect\subref{fig:real:nini:h97hw} initializations or \protect\subref{fig:real:time:h97hw} real time to target minimum for both algorithms.
Dotted lines indicate significance level controlling FWER at $0.05$.
Bottom plots show rate \protect\subref{fig:real:nini:h97hw} or time \protect\subref{fig:real:time:h97hw} ratios with 95\% confidence intervals on the log10 scale; no difference marked by the dotted line.
The $y$-axes are linked across facets in a row; 
$x$-axes are not linked.}
\label{fig:real:h97hw}
\end{figure*}

On the six real datasets introduced in Section~\ref{sec:real},
we find the proposed \hw\ algorithm is almost universally more accurate than the H97 algorithm per initialization, as measured by the proportion of initializations achieving a target minimum of the objective function~\eqref{eqn:objfun}.
To formally compare pairs of algorithms, 
we run them on the same initializations and count the number of runs where the first algorithm (H97) achieves the target minimum but the second (\hw) does not ($n_{10}$) and vice versa ($n_{01}$).
In Table~\ref{tab:real_results}, we report the average number of initializations between hits on the target for all three algorithms. 
The first four datasets achieve the minimum objective with ease, and the difference between H97 and the proposed algorithms is small, though significant.
\figref\ref{fig:real:h97hw}\,\subref{fig:real:nini:h97hw} shows details, including the probit-transformed left tail probabilities $\Pr(N_{10} \leq n_{10})$, which approach $\infty$ if the first algorithm is superior to the second or $-\infty$ if the second algorithm is superior.
Controlling the family-wise error rate (FWER) at $0.05$~\citep{Holm1979}, a null hypothesis of equal accuracy is strongly rejected in nearly all cases, and only the breast cancer dataset at $K=2$ found H97 significantly better than \hw{} (\figref\ref{fig:real:nini:h97hw}).
For all other datasets and choices of $K$, \hw\ was significantly more accurate per initialization than H97.

\begin{figure*}
  \centering
  \vspace{-1em}
  \mbox{\subfloat[Accuracy\label{fig:sim:k5:h97hw:nini:boxplot}]{\includegraphics[width=0.485\textwidth]{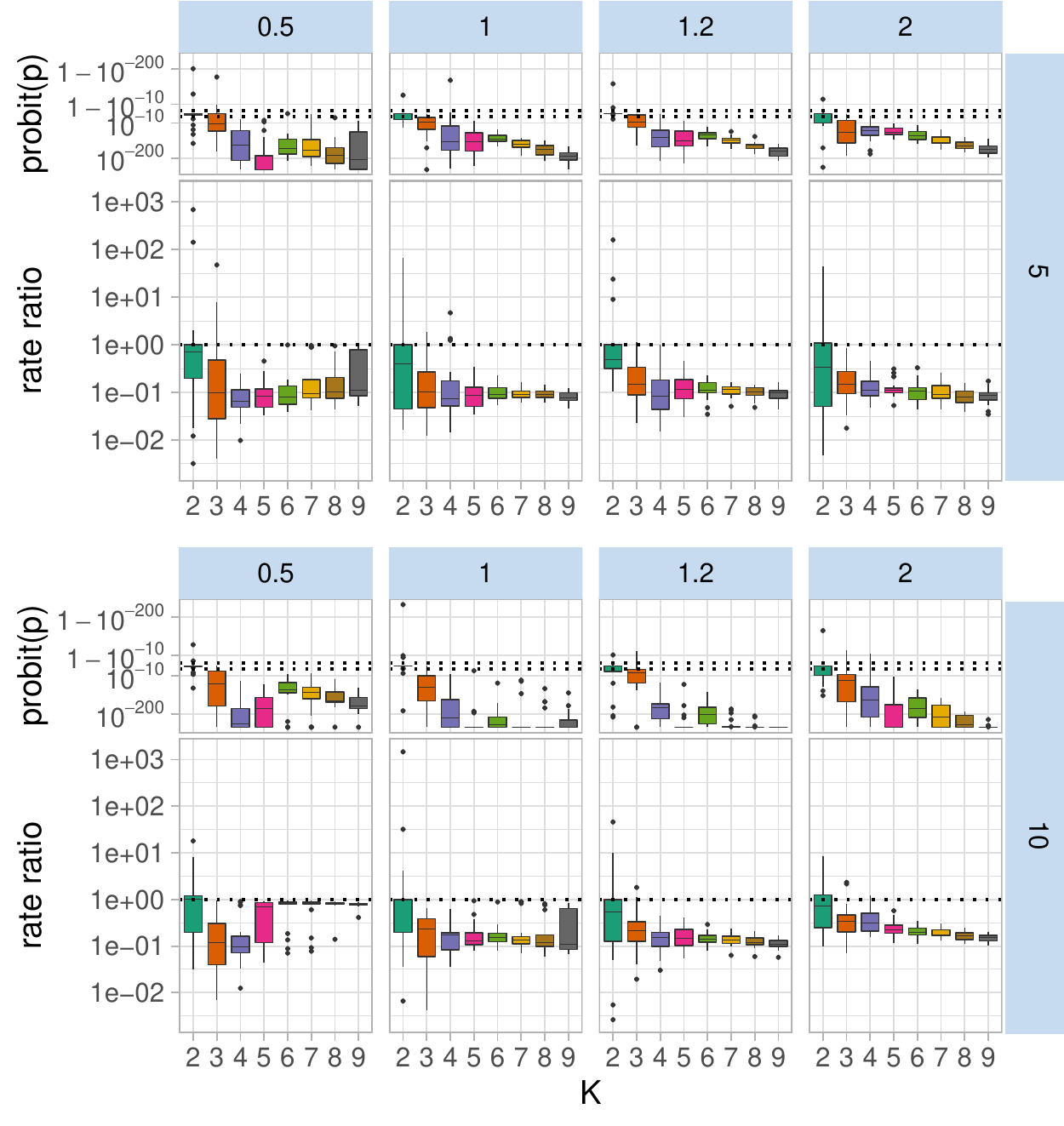}}
\hspace{1em}
    \subfloat[Timing\label{fig:sim:k5:h97hw:time:boxplot}]{\includegraphics[width=0.485\textwidth]{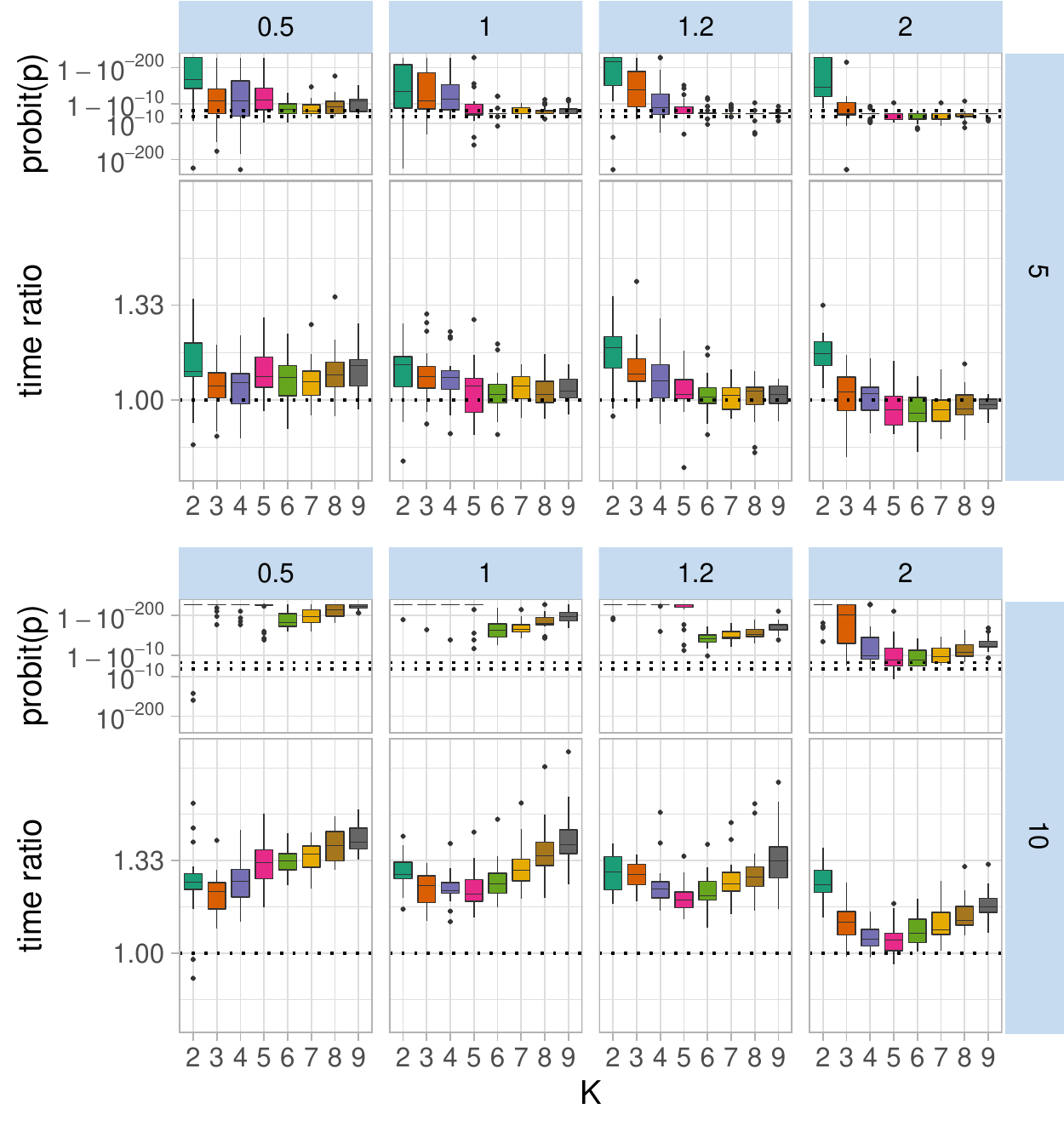}}}
\caption{Accuracy and timing of H97 vs.~\hw\ at $K_t=5$. In each pair of plots, boxplots summarize the probit-transformed, adjusted left-tail probabilities (upper plots) and log10-transformed ratios (lower plots) at each $K$ (horizontal position). 
The dotted horizontal lines in the upper plots demarcate the
acceptance region for the null hypothesis of no change when FWER is
0.05. Similar dotted lines in the bottom plots indicate no difference between the H97 and \hw\ algorithms.}
\label{fig:sim:k5:h97hw:boxplot}
\end{figure*}

The per iteration speed of H97 does not make up for fewer initializations arriving at the target; it takes more average time to achieve the target than the \hw\ algorithm.
To compare times, we collect data on the duration of each run, including initialization but excluding bookkeeping (\textit{e.g.}, for computing ARI).
We report average cumulative run time to achieve the target minimum for each algorithm using repeated initializations 
at the true $K_t$ in Table~\ref{tab:real_results}; other choices of $K$ and details are in \figref\ref{fig:real:h97hw}\,\subref{fig:real:time:h97hw}.
\hw{} is faster than H97 on all but the splice dataset (\figref\ref{fig:real:h97hw}\,\subref{fig:real:time:h97hw}).
H97 is a fraction faster than \hw{} on the breast cancer and soybean datasets for specific choices of $K$, but not on average across all $K$.
When H97 is faster than the \hw{} algorithm, it is rarely more than 30\% faster, whereas \hw{} is up to nine times faster than H97.
The speed advantage of the \hw{} algorithm increases with $K$, especially for $K$ larger than the true $K_t$.
Overall, \hw\ has a larger speed advantage on a broader range of datasets and conditions.

The theory of Section~\ref{sec:theory} offers no indication of the role of the quick transfer stage.
Empirically, we found \hw\ and \hwo\ were less likely to differ in performance from each other.
If they did, the effect size was typically smaller than the difference when comparing either method with H97.
For the senators and soybean datasets, \hw\ was more accurate than \hwo\ per initialization for intermediate values of $K$, while on the splice dataset, \hwo\ had a tiny, but significant, advantage over \hw\ (\figref\ref{fig:real:hwhwo}\,\subref{fig:real:nini:hwhwo}).
\hwo\ is significantly faster to the target minimum than \hw\ on all but the mushroom dataset (Table~\ref{tab:real_results} and \figref\ref{fig:real:hwhwo}\,\subref{fig:real:time:hwhwo}).
As a result, \hwo\ is faster than H97 on all but the soybean dataset at $K=3$ and the splice dataset at $K=4$ and $K=7$ (\figref\ref{fig:real:h97hwo:time}).

\subsubsection{Simulation Data}
Because we replicate each simulation condition $20$ times, it is possible to show trends across features of the data, such as dimension ($p$), number of clusters ($K_t$), and difficulty of clustering ($t$).
Whereas on real datasets, we compared methods across initializations,
for simulation data, here, we compare methods across replicates.

Just as for the real datasets, the \hw\ algorithm is significantly more accurate than the H97 algorithm as measured by the proportion of initializations achieving the target value of the objective function (\figref\ref{fig:sim:k5:h97hw:boxplot}\,\subref{fig:sim:k5:h97hw:nini:boxplot} for $K_t=5$, \figref\ref{fig:sim:k2:h97hw:boxplot}\,\subref{fig:sim:k2:h97hw:nini:boxplot} for $K_t=2$).
We plot the significance (transformed left tail probabilities, upper plots in each pair) and magnitude (log10 rate ratios, lower plots) of the difference in the targeting rates for the two algorithms.
The \hw\ algorithm is significantly more likely to achieve the target minimum than H97, except perhaps when true $K_t=5$ and estimation $K=2$.
The benefit of the \hw\ algorithm increases with estimation $K$ except when the clusters are most separated ($t=0.5$). 

\hw{} is almost always substantially faster than H97 (\figref\ref{fig:sim:k5:h97hw:boxplot}\,\subref{fig:sim:k5:h97hw:time:boxplot}).
As for the real data, the speed of H97 occasionally makes up the accuracy difference so that it achieves the target slightly faster than \hw{} under select conditions. 
At the specific setting $K_t=5$, $p=5$, $t=2$, and $K>4$, there is a small temporal advantage of H97 per initialization (average 0.11ms; interquartile range -0.16 -- 0.39ms), with H97 significantly better than \hw\ in $60$\% of the simulations.	
In contrast, the accuracy advantage of \hw\ at all settings with $p=10$ does translate into a speed advantage over H97 (average 11ms; interquartile rage 0.14 -- 8.9ms; significantly better in 87\% of simulations) that increases with $K$ so that by $K=9$, $100$\% of the differences are significant (average 44ms; interquartile range 11 -- 55ms).	
For the simulation with $K_t=2$, \hw\ is faster than H97 except for $K=2$ and good separation ($t=0.5$), although the advantage of \hw\ is often not significant at $K=3$ (\figref\ref{fig:sim:k2:h97hw:boxplot}\,\subref{fig:sim:k2:h97hw:time:boxplot}).
Overall, there appears to be little cost to using \hw\ and almost always an advantage. 

\section{Conclusions}
\label{conclusions}

We have devised a novel optimal transfer and quick transfer (\hw) $k$-modes algorithm, inspired by the $k$-means algorithm of \citet{hartiganandwong79}.
We prove our novel algorithm is capable of progressing to a lower value of the objective function even when competing algorithms would terminate.
On real and simulated data, we demonstrate that it is, in fact, more likely to achieve a better optimum given a single initialization.
Both the novel and original algorithms share $O(npK)$ run time per iteration~\citep{hartiganandwong79},
however, the proposed algorithm is computationally more expensive per iteration than the existing algorithm~\citep{huang97b}.
Nevertheless, \hw{} appears to scale better with complexity in the data, including more coordinates, more difficult clustering, and more clusters.
The previous algorithm can be faster under specialized conditions, such as when the clusters are well-separated and the assumed number of clusters is near the truth, but the absolute difference in time is typically small.
Especially considering that users generally choose the number of initializations, not the runtime, when applying $k$-modes, the more accurate and almost always faster \hw\ algorithm should be preferred.

The timing results are sensitive to the implementation, platform used for testing (Intel(R) Xeon(R) CPU E3-1241 v3 \@ 3.50GHz), and compiler (gcc).
In our hands, the alternative algorithm \hwo{} that dispenses with the quick transfer stage ends up as the fastest algorithm to the minimum objective.
In polishing the algorithm description for publication, we also identified a possible improvement for the optimal transfer stage.
Currently, when a transfer happens, the second closest mode is set to the previous mode, but the algorithm has information about closer modes if they exist.
There may also be further enhancements possible for the H97 algorithm.
Thus, tweaks to the algorithms and more efficient coding may change the temporal differences between the algorithms, but the higher per-initialization accuracy of \hw{} and \hwo{} over H97 is proven.

Two sets of tie-breaking rules, chosen largely for their algorithmic convenience, are used in all algorithms.
First, we arbitrarily order categories to distinguish major and minor modes when there are tied category counts within a cluster.
Second, we use the cluster order to determine how to assign observations to clusters when two or more modes are equally distant.
These rules guarantee the same solution given the same initial modes in the same order, but may yield different clusters when the same initial modes are given in different order.
To guarantee the same solution despite initial mode ordering, the second rule could be modified to transfer to the first cluster ordered by the categories at each coordinate.
However, neither solution reflects the true uncertainty in the data.
It would be preferable to code algorithms that randomize both the category and mode orderings and then summarize the observed ambiguity in modes and partitions across the multiple solutions achieving the same minimum objective value.
In reality, most algorithms will either report the first or last solution to obtain the minimum objective value.

The $k$-modes algorithm is not the only clustering method for categorical data.
Most approaches transform the data so that numeric clustering methods can apply~\citep{ralambondrainy95,Carroll1986,Chaturvedi2001}.
Such approaches include defining a distance~\citep{gowdaanddiday91,Cao2012,Ienco2012,Jia2016,Kim2012,Le2005,Ng2007,Zhou2017a} or similarity~\citep{Goodall1966b,gower71,Gowda1992,Ahmad2007,Sangam2015,Santos2015} or link~\citep{Guha2000} for categorical data and using the result in an appropriate clustering algorithm.
Others optimize, often approximately, criteria defined for categorical data~\citep{Lazarsfeld1950,Goodman1974,Clogg1984,huang97b,Barbara2002,Ganti1999,Gibson2000,Ng2002,He2002a,Cesario2007,He2008,Heloulou2017,Andritsos2004,Li2014d,Park2015,Parmar2007,Thakur2017,Wong1979,Zhang2006a}.
Each of these methods and $k$-modes assume, 
 either explicitly or implicitly, some generative model, and while
 there have been some comparisons of their
 performance~\citep{Anderlucci2014,Bai2015}, there is no doubt that the $k$-modes objective function is not always the best optimization criteria for categorical data.
In many cases, where the generative model is unknown, it may be preferable to employ ensemble methods for combining results from multiple categorical data clustering methods~\citep{He2005,Iam-On2012,Saha2015a,Amiri2018}.
Ensemble methods rely on algorithmic efficiency, so when $k$-modes is
included in the ensemble, it will be useful to employ the fastest
possible version of the algorithm, such as \hw{} and \hwo{}. Further,
the new algorithms could be incorporated in syncytial clustering algorithms for
categorical data using Generalized or Gaussianized     Distributional
    Transforms~\citep{ruschendorf13,daietal19} 
    and copula models~\citep{nelsen06}, as outlined
in \citet{almodovarriveraandmaitra20}.
In conclusion, we have contributed a novel and efficient $k$-modes algorithm that can be combined with many additional ideas and solutions to solve remaining problems in categorical data clustering.
\appendix

\begin{figure}[h]
\subfloat[\label{fig:real:nini:hwhwo}]{\includegraphics[width=0.47\textwidth]{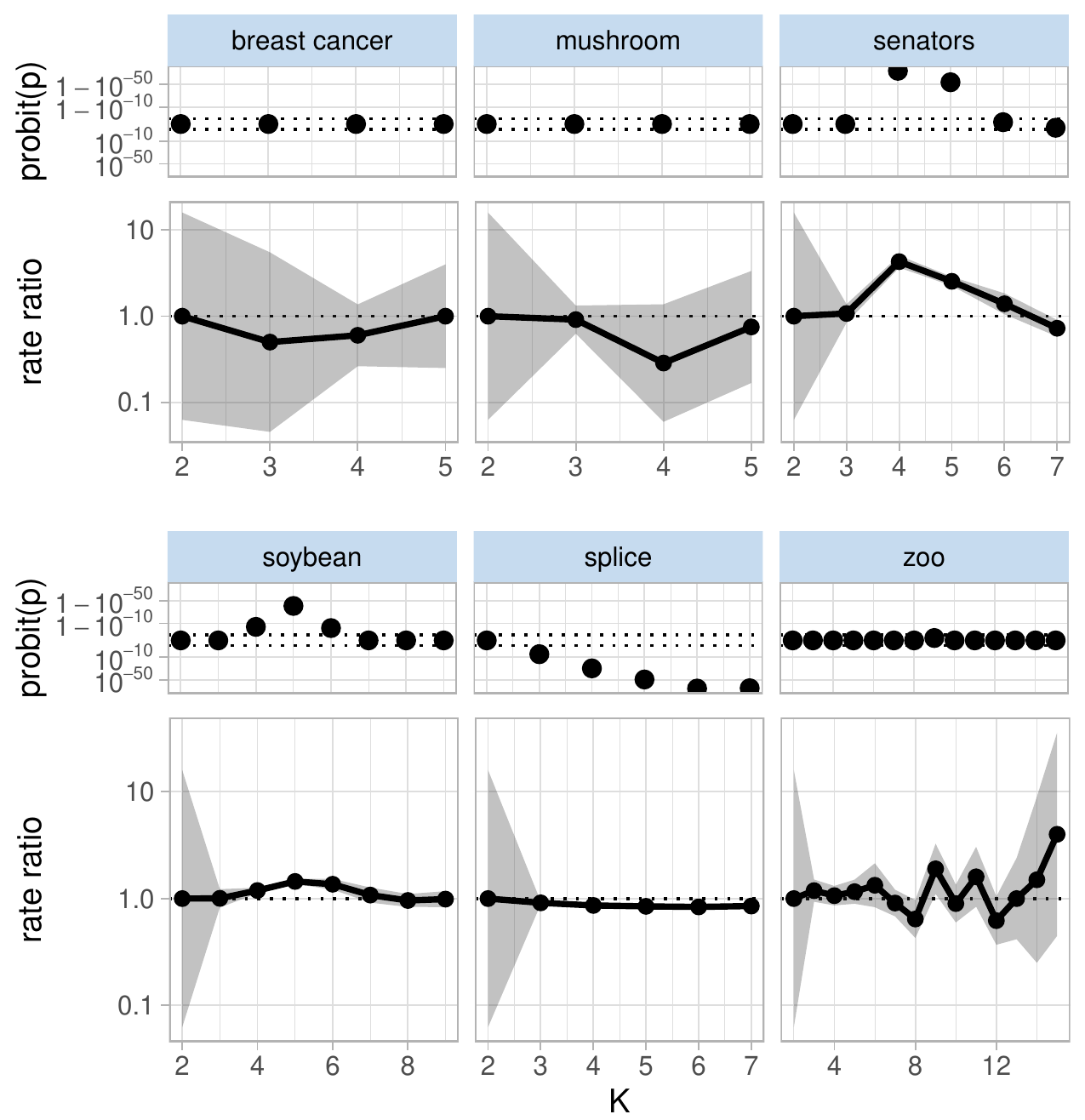}}
\hspace{2em}
\subfloat[\label{fig:real:time:hwhwo}]{\includegraphics[width=0.47\textwidth]{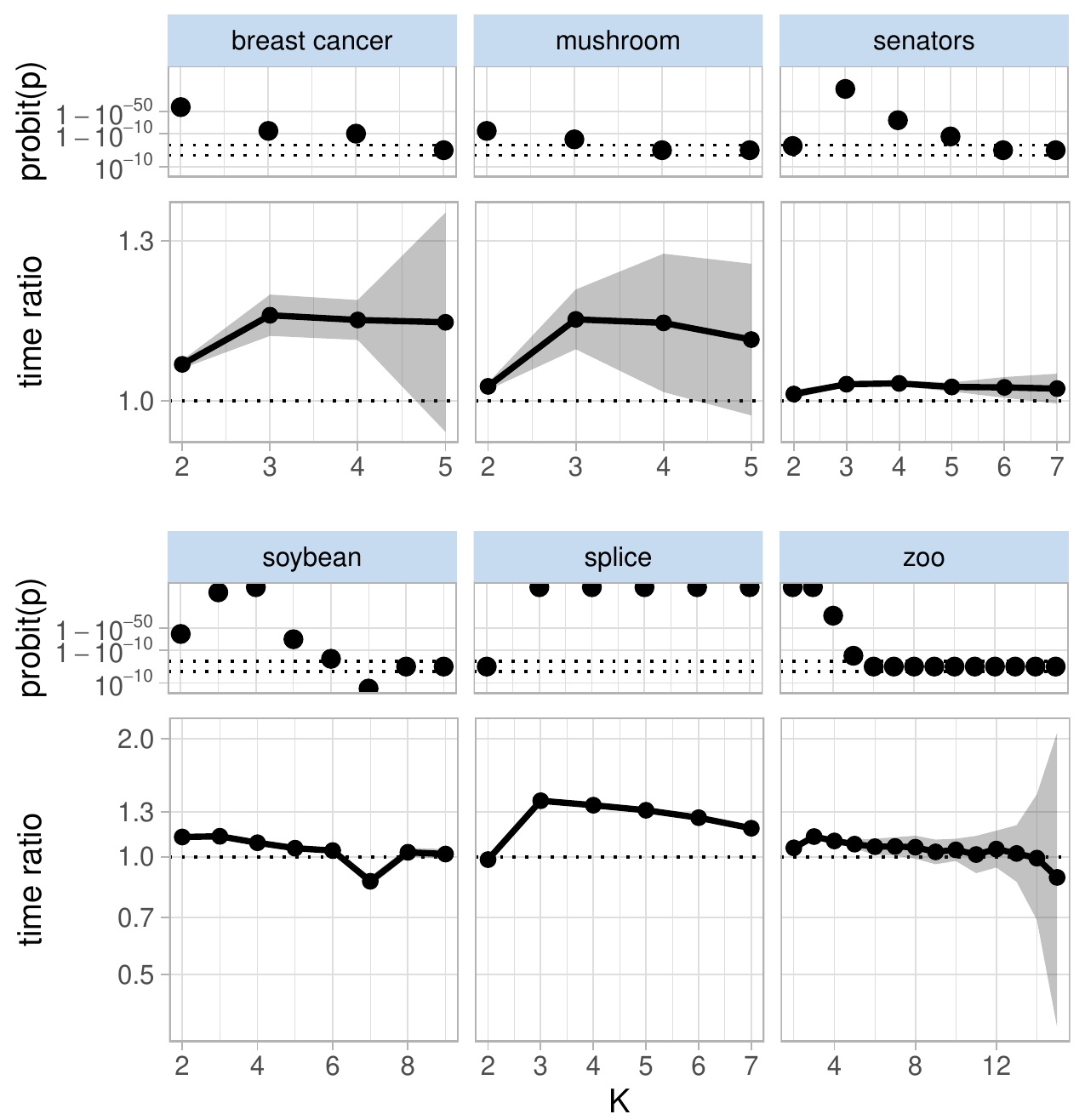}}
\caption{
Comparison of \hw\ and \hwo\ on achieving target minimum.
See \figref\ref{fig:real:h97hw} legend for further explanation.
\label{fig:real:hwhwo}
}
\end{figure}

\begin{figure}[t]
\subfloat[Accuracy\label{fig:sim:k2:h97hw:nini:boxplot}]{\includegraphics[width=0.47\textwidth]{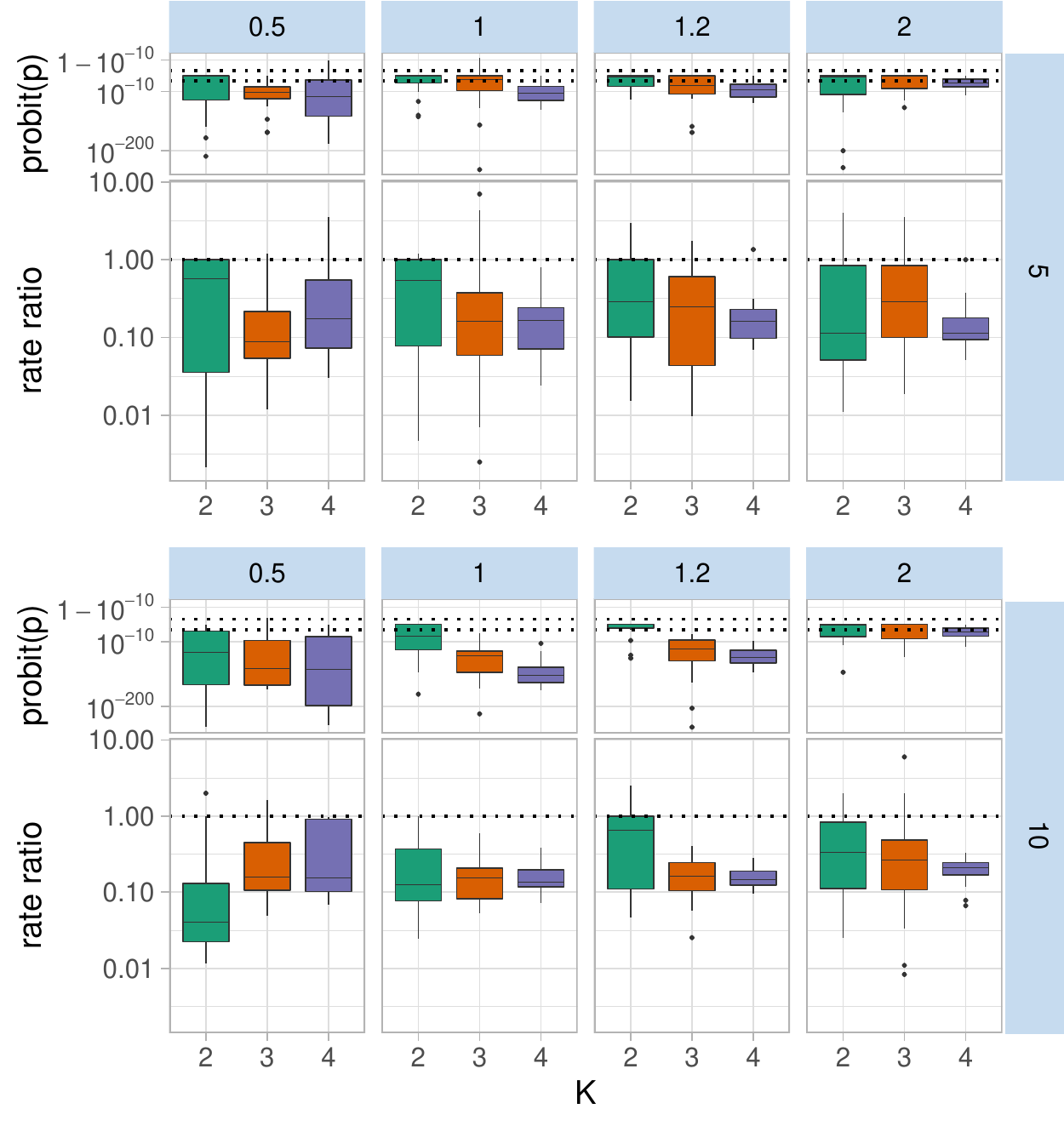}}
\hspace{2em}
\subfloat[Timing\label{fig:sim:k2:h97hw:time:boxplot}]{\includegraphics[width=0.47\textwidth]{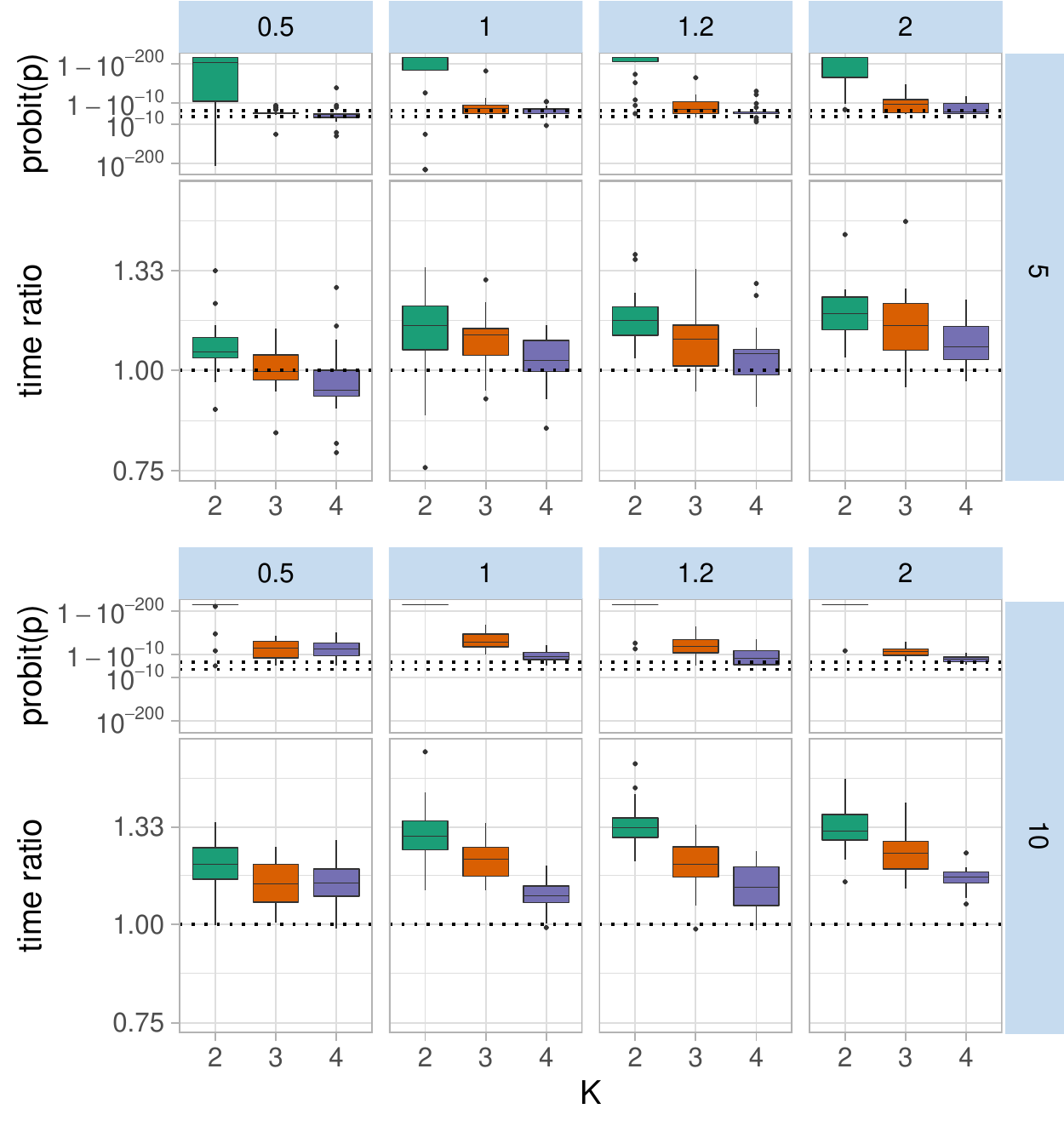}}
\caption{Accuracy and timing of H97 vs.~\hw\ for $K_t=2$.\label{fig:sim:k2:h97hw:boxplot}
See \figref\ref{fig:sim:k5:h97hw:boxplot} legend for further explanation.}
\end{figure}

\begin{figure}[t]
\subfloat[Accuracy\label{fig:sim:k5:hwhwo:nini:boxplot}]{\includegraphics[width=0.47\textwidth]{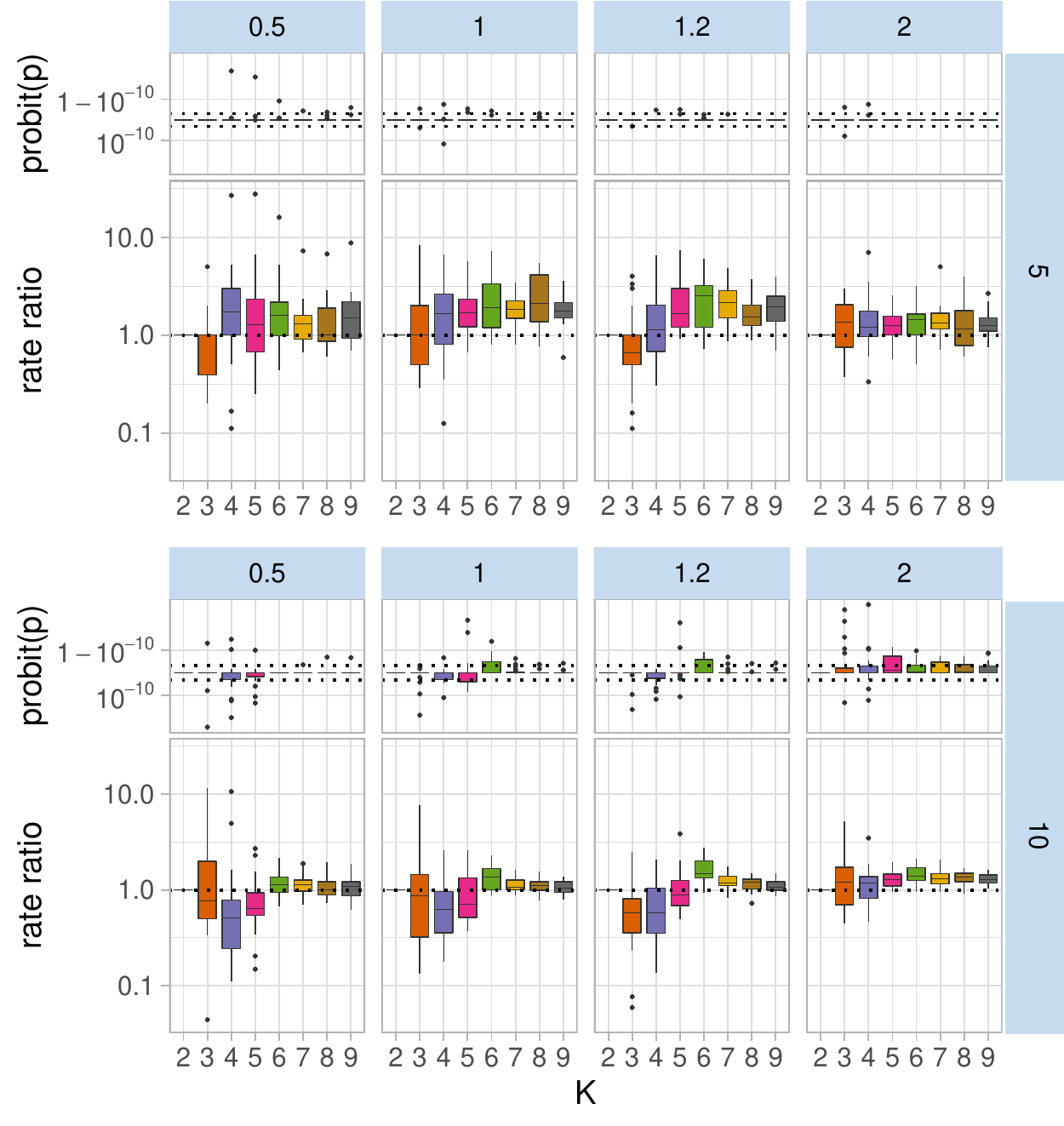}}
\hspace{2em}
\subfloat[Timing\label{fig:sim:k5:hwhwo:time:boxplot}]{\includegraphics[width=0.47\textwidth]{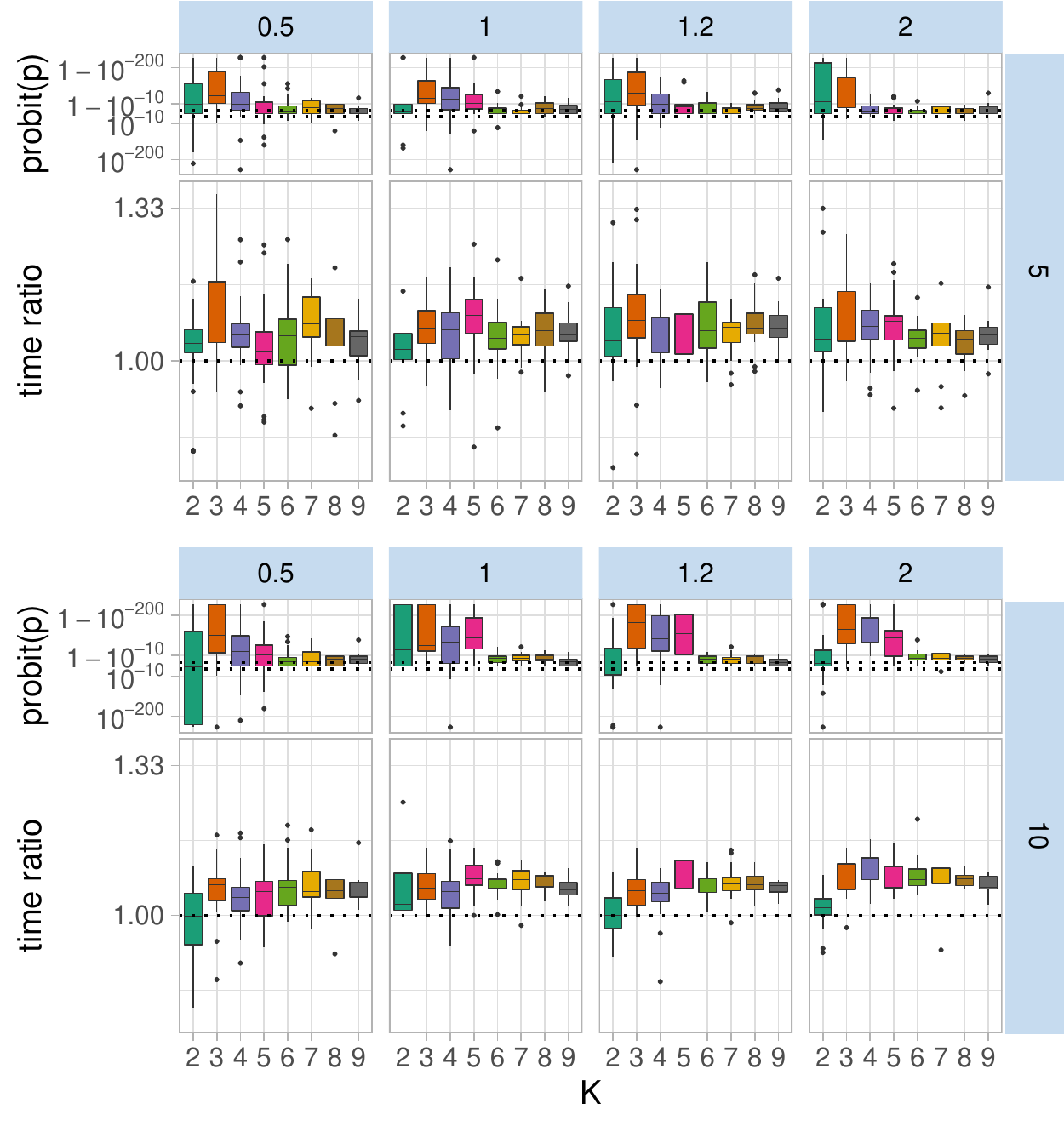}}
\caption{Accuracy and timing of \hw\ vs.~\hwo\ for $K_t=5$.\label{fig:sim:k5:hwhwo:boxplot}
See \figref\ref{fig:sim:k5:h97hw:boxplot} legend for further explanation.}
\end{figure}

\begin{figure}[t]
\subfloat[Accuracy\label{fig:sim:k2:hwhwo:nini:boxplot}]{\includegraphics[width=0.47\textwidth]{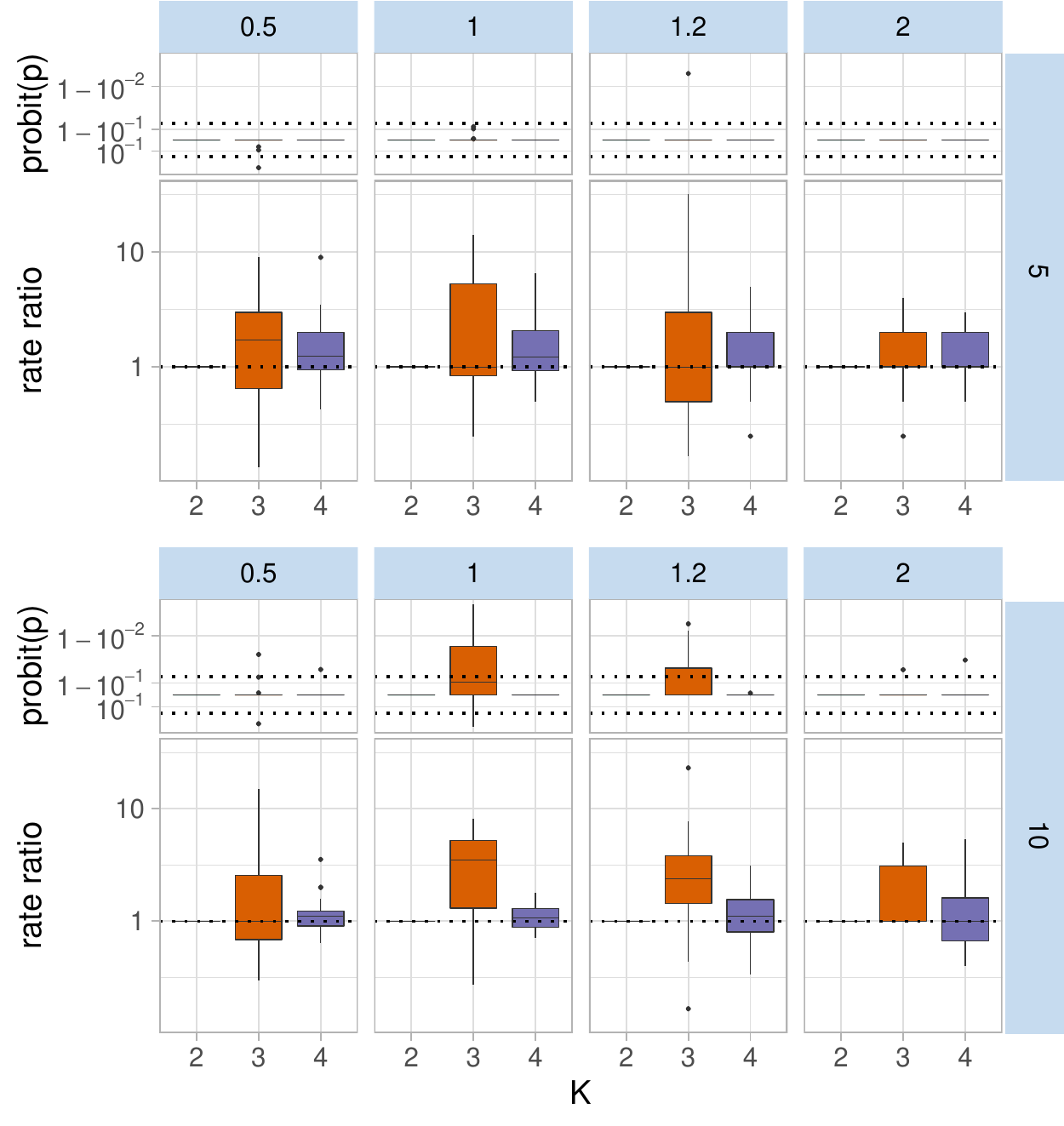}}
\hspace{2em}
\subfloat[Timing\label{fig:sim:k2:hwhwo:time:boxplot}]{\includegraphics[width=0.47\textwidth]{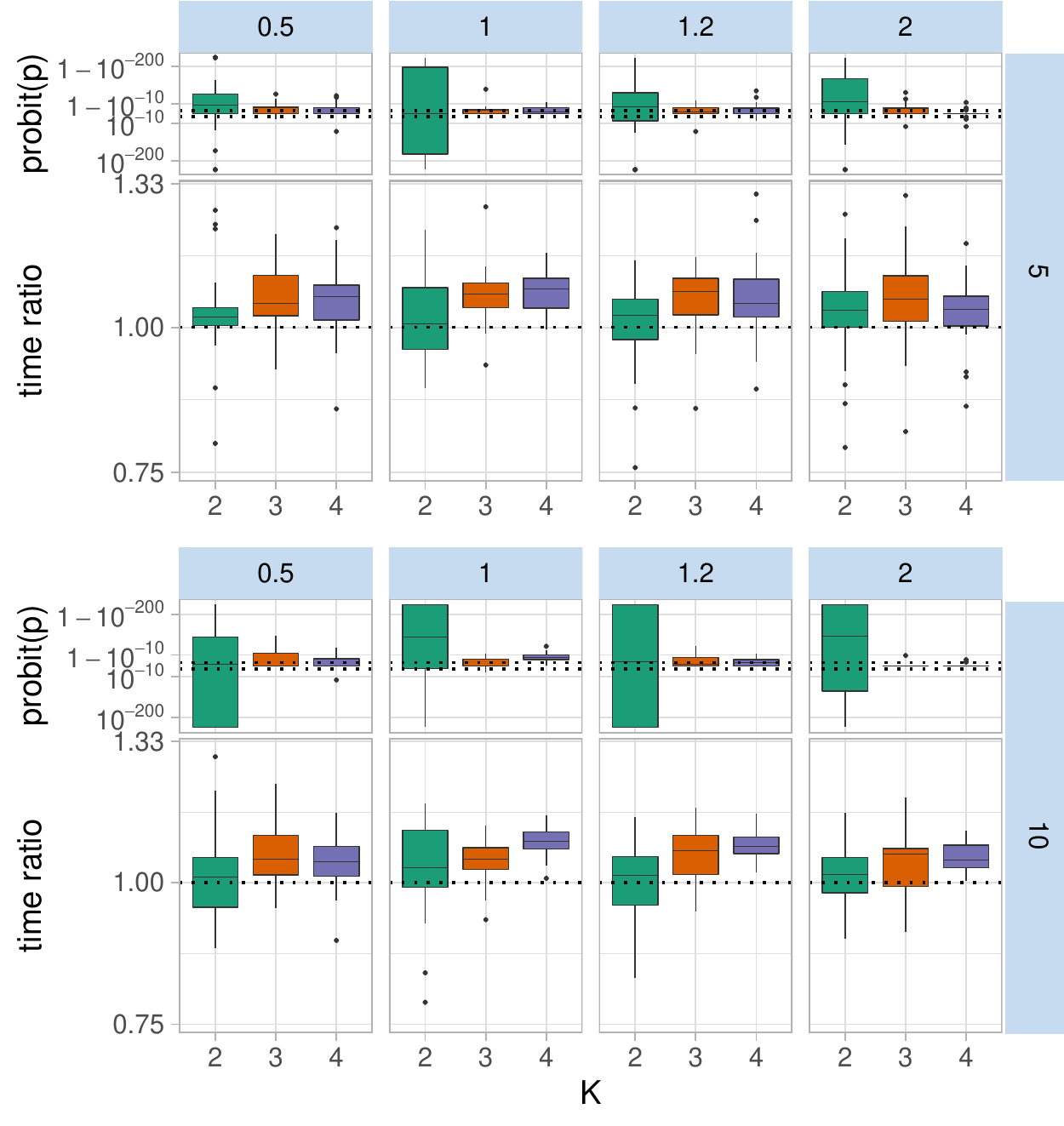}}
\caption{Accuracy and timing of \hw\ vs.~\hwo\ for $K_t=2$.\label{fig:sim:k2:hwhwo:boxplot}
See \figref\ref{fig:sim:k5:h97hw:boxplot} legend for further explanation.}
\end{figure}

Here we include supplementary figures for additional analyses that are mentioned but not displayed in the main text.
The figures are ordered as they are referenced in the main text.
The main text focuses on the comparison of algorithms H97 and \hw{}, but here we show there are smaller, but sometimes significant differences between the \hw{} and \hwo{} algorithms.
Similarly, the main text focuses on results for simulation data with $K_t=5$, but similar patterns are found for the $K_t=2$ simulations.
In Figs.~\ref{fig:real:hwhwo},~\ref{fig:sim:k5:hwhwo:boxplot} and~\ref{fig:sim:k2:hwhwo:boxplot}, we compare \hw{} and \hwo{} in terms of the number and timing of initializations reaching the target for the real data, $K_t=5$ simulation data, and $K_t=2$ simulation data.
Fig~\ref{fig:real:h97hwo:time} compares time to target for directly comparing H97 and \hwo{} on the real data, showing that \hwo{} is even more likely than \hw{} to beat H97 to the target.
\figref\ref{fig:sim:k2:h97hw:boxplot} shows the accuracy and timing results for comparing algorithms H97 and \hw{} for $K_t=2$ simulation data.

\begin{figure}[h]
\centering
\includegraphics[width=0.47\textwidth]{fig.s1/real_pt_h97-hwo_time}
\caption{
Comparison of H97 and \hwo\ on time to achieve target minimum.
See \figref\ref{fig:real:h97hw} legend for further explanation.
\label{fig:real:h97hwo:time}
}
\end{figure}

\bibliographystyle{IEEEtran}
\bibliography{references}

\end{document}